\documentclass{aa}  

\makeatletter
\renewcommand*\aa@pageof{, page \thepage{} of \pageref*{LastPage}}
\makeatother
\usepackage{lscape}
\usepackage{supertabular}
\usepackage{longtable}

\usepackage{graphicx}
\usepackage{txfonts}
\usepackage{xcolor}
\usepackage{multirow}
\usepackage{hyperref}
\hypersetup{
  colorlinks=true,
    linkcolor=blue,
    citecolor=blue,
    urlcolor=blue}

\bibpunct{(}{)}{;}{a}{}{,} 

\begin{document}

   \title{Is convective turbulence the only exciting mechanism of global 
\textit{p}\,modes in the Sun?}

   \author{E. Panetier\inst{1} \and
           R. A. García\inst{2} \and
           S. N. Breton\inst{3} \and
           A. Jiménez\inst{4, 5} \and 
           T. Foglizzo\inst{2}
          }

   \institute{Université Paris Cité, Université Paris-Saclay, CEA, CNRS, AIM, 91191, Gif-sur-Yvette, France\\
              \email{eva.panetier@cea.fr}
         \and
             Université Paris-Saclay, Université Paris Cité, CEA, CNRS, AIM, 91191, Gif-sur-Yvette, France
         \and
             INAF – Osservatorio Astrofisico di Catania, Via S. Sofia, 78, 95123 Catania, Italy
         \and
             Instituto de Astrofísica de Canarias
             (IAC), 38205 La Laguna, Tenerife, Spain
         \and
             Universidad de La Laguna (ULL), Departamento de Astrofísica, 38206 La Laguna, Tenerife, Spain
             }

   \date{}

 
  \abstract
   {In solar-like oscillators, acoustic waves are excited by turbulent motion in the convective envelope and propagate inward, generating a variety of standing pressure modes (\textit{p}\,modes).
   When combining together the power of several solar acoustic modes, an excess not compatible with pure stochastic excitation was found in some studies. This could be the signature of a second mode excitation source.}
   {With over 27 years of helioseismic data from the Sun as a star observations by the Solar and Heliospheric Observatory (SoHO), we aim to study the variation in mode energy over this period, covering solar Cycles 23, 24, and the beginning of Cycle 25. In particular, we focus on the possible sources of high peaks in the mode-energy time series, i.e. instrumental problems or other exciting mechanisms, such as flares or Coronal Mass Ejections (CMEs).}
   {We reconstruct the energy time series for each mode, resulting in 36 time series with a sampling time of 1.45 days. By combining the small-time-scale variations in energy for several low-degree modes ($\ell\leq2$) in the 2090-3710\,$\mu$Hz range, we study the correlation between the modes and their compatibility with the hypothesis that modes are only stochastically excited by convection.}
   {The observed excitation rate significantly deviates from what would be expected in the case of a purely stochastic excitation. Our results indicate that this energy excess cannot be only attributed to instrumental effects and does not exhibit a cyclic variation. Although high-energy excesses are occasionally associated with observations of flares and/or CMEs, no consistent pattern could be identified. The excitation is slightly more frequent for modes probing the upper layer of the convective zone. Furthermore, the energy supply rate seems to vary over time with the mean value following a modulation that can match the Quasi-Biennial Oscillation (QBO) observed in other solar indicators, and the variance being anti-correlated with the cycle. 
   }
   {}
   
   \keywords{ Sun: helioseismology -- Sun: activity -- Sun: interior -- Methods: data analysis
               }

   \maketitle
%

\section{Introduction}
\label{sec:intro}

Observing stellar oscillations provides constraints on the internal properties of the stars. Variable stars are found across the entire Hertzsprung-Russell diagram, and various mechanisms are capable of exciting their pulsations, depending on their fundamental properties and structure \citep[e.g. ][ and references therein]{christensen-dalsgaard_2014, aerts_2024}. 

Solar-like stars are composed of an internal radiative zone surrounded by an external convective envelope. In these stars, convection operates in a regime of fully developed turbulence, characterised by extremely high Reynolds numbers ($Re >>  10^6$). This turbulent flow transfers energy from large convective scales down to the small dissipative scales \citep{kolmogorov_1941}. In this complex dynamical regime, the turbulent motions of the stellar plasma, along with fluctuations of thermodynamic quantities, randomly inject energy into the star’s oscillation eigenmode — in particular, the acoustic (or \textit{p}) modes — through a mechanism known as stochastic excitation \citep[e.g.][]{goldreich_1977, samadi_2001, belkacem_2008}.
This process leads to the continuous excitation of several oscillation modes, which are simultaneously driven and damped \citep{belkacem_2012}  — both processes occurring on different characteristic timescales \citep{samadi_2015}. Consequently, the power spectra of these modes exhibit Lorentzian profiles, with amplitudes determined by the balance between excitation and damping.

Moreover, solar-like stars also experience magnetic field variations, driven by a dynamo mechanism \citep{mathur_2014}. The solar magnetic field evolves following an 11-years cycle (the Schwabe cycle) with a change in polarity at the end of each cycle. The modes frequency and amplitude variations in relation with the solar magnetic cycle has been extensively studied \citep[e.g.][]{garcia_2013, salabert_2015}, as well as modes parameter variations for magnetically active solar-like stars other than the Sun \citep[e.g.][]{regulo_2016, santos_2019}. Frequencies increase with the magnetic field strength whereas amplitudes decrease. \citet{bessila_2024} derived the formalism describing the excitation of the \textit{p}\,modes by convective motions in presence of a magnetic field, and demonstrated that the mode amplitude decreases with the magnetic field strength.
Furthermore, previous studies have measured a constant energy supply rate over time for low-degree modes, attributing this constancy to a balance between damping and excitation processes \citep{chaplin_2000, jimenez-reyes_2004}. However, for the first time, \citet{kiefer_2021} observed that the energy supply rate varies over the cycle for intermediate-degree modes ($2 < \ell < 150$), this variation being anti-correlated with the solar cycle. They further suggest that modes with frequencies below 3000\,$\mu$Hz are more strongly influenced by this variation, as well as modes with degrees greater than $\ell \sim 20$.

After having studied the correlation between modes energies with data from the InterPlanetary Helioseismology by IRradiance measurements \citep[IPHIR, ][]{frohlich_1988} and the 310 first days of data from the Global Oscillation at Low Frequencies \citep[GOLF, ][]{gabriel_1995} of the Solar and Heliospheric Observatory \citep[SoHO, ][]{domingo_1995}, \citet{baudin_1996} and \citet[further cited F+98]{foglizzo_1998} found a correlation between several low degree modes. F+98 even noticed that several modes were excited at the same time during the end of 1998. Because this is not expected if modes are only stochastically excited by convection, they suggested the existence of a second source of mode excitation.
The possibility of high-energy flares, specifically X-class flares with a flux greater than $10^{-4}$~W\,m$^{-2}$, and coronal mass ejections (CMEs) injecting energy into modes has been investigated by \cite{foglizzo_1998}. The typical size of flares observed from the Earth is on the order of a few arcminutes, suggesting that they have the potential to excite over $10^4$ modes. However, assuming an equal distribution of energy, their intensity is insufficient to supply the necessary energy for the excitation of each acoustic mode. Nevertheless, the typical latitude distance between the foot points of the outer loop of the CMEs is about 45 degrees. Considering this as the CME's typical size, they should be able to excite less than a hundred low-degree modes. Based on an estimation of the CMEs' kinetic energy, \cite{foglizzo_1998} suggested that they may transfer sufficient energy to excite these modes. With more than 28 years of continuous observations from the Solar and Heliospheric Observatory (SoHO) available now, covering solar magnetic Cycles 23, 24, and approximately the first half of Cycle 25, the relationship between the energy of these modes and solar magnetic variations must be re-investigated, and compared to flare and CME occurrences.

Heliospheric data are described in Sect.~\ref{sec:observations} and the data processing to reconstruct an energy time series for each mode is described in Sect.~\ref{sec:data_processing}. Section~\ref{sec:analysis} explains our statistical analysis on the energy variation of the modes, which is discussed in Sect.~\ref{sec:solutions}.

\section{Helioseismic time series}
\label{sec:observations}

The Solar and Heliospheric Observatory \citep[SoHO,][]{domingo_1995} was launched in December 1995 embarking onboard three instruments for helioseismology: the Michelson Doppler Imager \citep[MDI,][]{scherrer_1995} of the Solar Oscillation Imager (SOI), the Global Oscillation at Low Frequencies \citep[GOLF,][]{gabriel_1995} spectrometer, and the Variability of solar IRradiance and Gravity Oscillations \citep[VIRGO,][]{frohlich_1995}. Measurements of GOLF and VIRGO SunPhotoMeters (SPMs) are integrated over the solar disk and can therefore be compared with other stars observations. GOLF observes in its single wing configuration: i.e. it measures mainly the blue wing of the sodium doublet at 589.6 and 589.0\,nm since April 11$^{\mathrm{th}}$, 1996, although it was switched to the red-wing configuration after the SoHO recovery mission on September 1998 to November 18$^{\mathrm{th}}$, 2002 \citep[see][]{garcia_2005}. From these measurements, a proxy of the Doppler velocity is produced as shown in \citet{palle_1999}. The cadence of the observations is 5\,s but the time series are then calibrated and resampled to multiples of 20\,s. VIRGO/SPMs measures the sunlight irradiance integrated over the solar disk since January 23$^{\mathrm{rd}}$, 1996, with a 60\,s cadence, in three wavelengths of the optical band: 402\,nm (blue channel), 500\,nm (green channel), and 862\,nm (red channel).
In this work, an asteroseismic proxy was built combining green and red channels from VIRGO/SPMs measurements, as this combination is the most similar to the \textit{Kepler} passband \citep{basri_2010}, allowing more straightforward comparisons with other stars.
The calibration of these data is described in Appendix~\ref{app:calibration_VIRGO}. GOLF and VIRGO/SPMs measurements cover a period from the minimum between solar magnetic Cycles 22 and 23 to the ongoing maximum of Cycle 25.


\section{Data processing: temporal evolution of a single mode}
\label{sec:data_processing}

\begin{figure}
   \centering
   \includegraphics[width=0.5\textwidth,trim={0 0 0 0},clip]{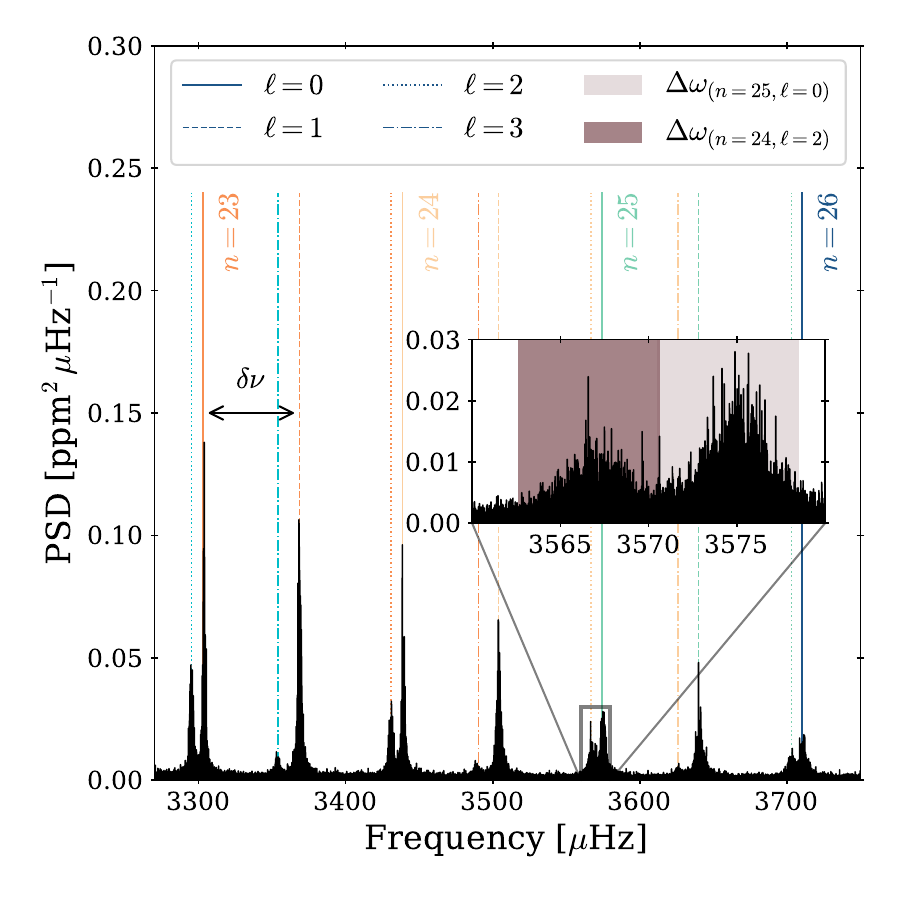}
      \caption{PSD of the VIRGO/SPMs light curve (from 23.01.1996 to 23.11.2023). Higher frequencies are displayed between 3270 and 3750\,$\mu$Hz showing the large frequency separation ($\delta\nu$) noted between modes $\ell=0$ and $\ell=1$ for $n=23$. Modes are marked by vertical lines with different colours: $n=22$ in light blue, $n=23$ in orange, $n=24$ in yellow, $n=25$ in cyan and $n=26$ in dark blue). $\ell=0$ are in continuous, $\ell=1$ in dashed, $\ell=2$ in dotted and $\ell=3$ in dot-dashed lines. In the subplot: the window $\Delta\omega = 8\,\mu$Hz used to reconstruct modes energy time series is shown around modes ($n=25$, $\ell=0$) in light brown and ($n=24$, $\ell=2$) in dark brown.}
         \label{fig:modes_identification}
   \end{figure}

The observed time series depend on many contributions such as rotation, granulation, solar activity, oscillation modes, photon noise; each of them having a contribution at a distinctive frequency range. The energy contribution from a mode can be isolated and extracted, following the procedure first described by F+98. By isolating the contribution of each mode by filtering in the Fourier domain and computing its inverse Fourier transform, we can reconstruct its energy time series. Following the analogy with a simple harmonic oscillator of eigenfrequency $\omega_0$, F+98 showed that the surface energy per unit of mass $\varepsilon_{\mathrm{n,\ell}}$ of a mode can be approximated by: 
\begin{equation}
    \varepsilon_{\mathrm{n,\ell}}(t) = \frac{E(t)}{M} = 2\left|f_v(t)\right|^2 \left\{1 + \mathcal{O} \left(\frac{\Delta\omega}{ \omega_0}\right)  \\\right\}\,,
    \label{eq:energy}
\end{equation}where $f_v(t)$ is the inverse Fourier transform of the oscillatory velocity and $M$ is the mode mass. If $\hat{v}(\omega)$ is the single mode-filtered Fourier transform of the original time series, according to the window of size $\Delta\omega$, $f_v(t)$ can be written:
\begin{equation}
    f_v(t) = \int^{+\frac{\Delta\omega}{2}}_{-\frac{\Delta\omega}{2}}{\hat{v}(\omega_0+\omega)e^{i\omega t}\,\mathrm{d}\omega}\,.
    \label{eq:fvt}
\end{equation} 

The steps of the reconstruction method are summarised as follows:
\begin{enumerate}
    \item Computation of the Fourier Transform of the time series.
    \item Selection of the central frequency of a mode in the Fourier domain according to the window $\Delta\omega$ centred on the central mode frequency. The filtering is performed by setting all frequencies outside this window to zero. In Fig.~\ref{fig:modes_identification}, the PSD of $p$\,modes between 2400\,$\mu$Hz and 2900\,$\mu$Hz from GOLF observations is shown. Two windows of 8$\,\mu$Hz, centred on the middle frequency of the modes ($n=24$, $\ell=2$) and ($n=25$, $\ell=0$) are illustrated.
    \item Translation of the window towards central frequency $\omega_0 = 0$.
    \item Computation of the inverse Fourier transform of it according to Eq.~\eqref{eq:fvt}.
    \item The energy time series of the mode $\varepsilon_{\mathrm{n,\ell}}(t)$ is given by Eq.~\eqref{eq:energy}.
\end{enumerate}
   
\begin{figure}
   \centering
   \includegraphics[width=0.49\textwidth,trim={0 0 0 0},clip]{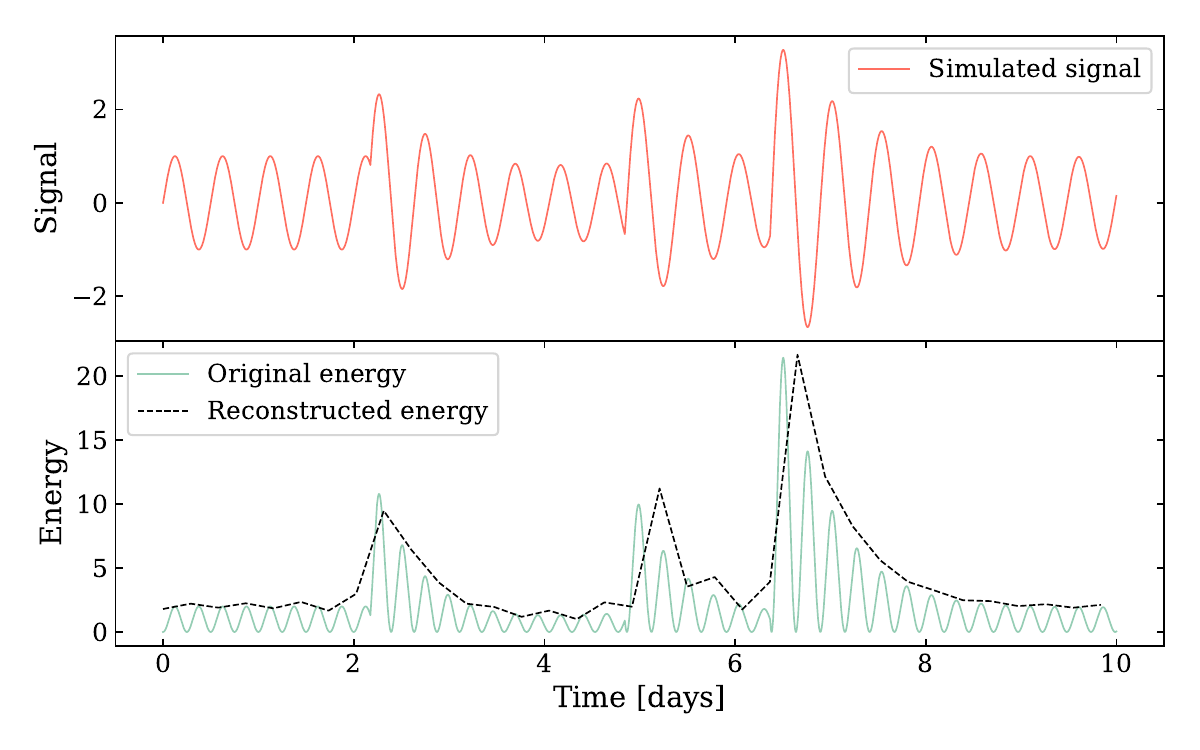}
      \caption{Top panel: Simulated sinusoidal signal $S(t)$ with a period of 0.5\,days and three excitations randomly distributed in time. It is shown for a time series of 10\,days sampled at 60\,seconds. Excitations are exponentially decaying. Bottom panel: energy $E(t)=2|S(t)|^2$ corresponding to the simulated signal is shown in grey and the reconstructed energy is shown in dashed black.}
         \label{fig:method_on_simu}
\end{figure}

To validate our implementation, we applied this process on a 1/2\,day-period simulated signal of a quasi-periodic oscillation with three random excitations. Results of the reconstruction with a window of $40\,\mu$Hz are displayed in Fig.~\ref{fig:method_on_simu} for a time series of 10\,days sampled at 60\,seconds. Periods scales of this simulated signal were deliberately chosen as different from what is expected from the modes for more readability. Comparing $\varepsilon(t)$ with the reconstructed energy clearly demonstrates that this method retrieves the envelope of the energy time series, which is essential for determining the excitation times.
We validated that it respects Parseval's theorem, comparing the integrals of the squares of the filtered signal in both the Fourier domain and the time domain. In Fig.~\ref{fig:method_on_simu} the time resolution difference we get between the original signal and the reconstructed one is also shown. 
As discussed by F+98, the time resolution $\delta t$ of the energy time series is related to the size of the filtering window $\Delta\omega$:
\begin{equation}
    \delta t = \frac{1}{\Delta \omega}\,.
\end{equation}
It sets a limit of binning for the reconstructed signal, ruled by both the window size and the resolution in the Fourier domain $1/T$, where $T$ is the total observation length: the number of points in the reconstructed signal is $p = T\,\Delta\omega$. 
The larger the window size the better the resulting resolution of the reconstructed time series. Moreover, $\Delta \omega$ should be large enough to include the total power of the mode peak but small enough for windows around nearby modes not to overlap each other. For example, a maximum window of $8\,\mu$Hz is illustrated on Fig.~\ref{fig:modes_identification} around high frequencies modes ($n = 25$, $\ell = 0$) and ($n = 24$, $\ell = 2$).
As higher frequency modes are broader, it is harder to set a window size for higher radial order modes, but in this case, it is possible to reconstruct the energy time series of several successive modes at a time (e.g. $\ell=0$ and $\ell=2$ or $\ell=1$ and $\ell=3$ modes), keeping in mind that several modes contributed to the resulting time series.
For the solar case, the time series are sufficiently resolved in the frequency domain, allowing us to consider one mode at a time.

\section{Analysis of the energy time series}
\label{sec:analysis}

We begin by describing the reconstruction of mode energy for the VIRGO/SPMs data. Subsequently, statistical tests are presented to evaluate the stochastic and non-stochastic behaviour of the mode energy.

\subsection{Reconstruction of the VIRGO/SPMs energy time series}
\label{subsec:reconst_VIRGO}

\begin{figure}
   \centering
   \includegraphics[width = 0.5\textwidth,trim={0.5cm 0 0 0},clip]{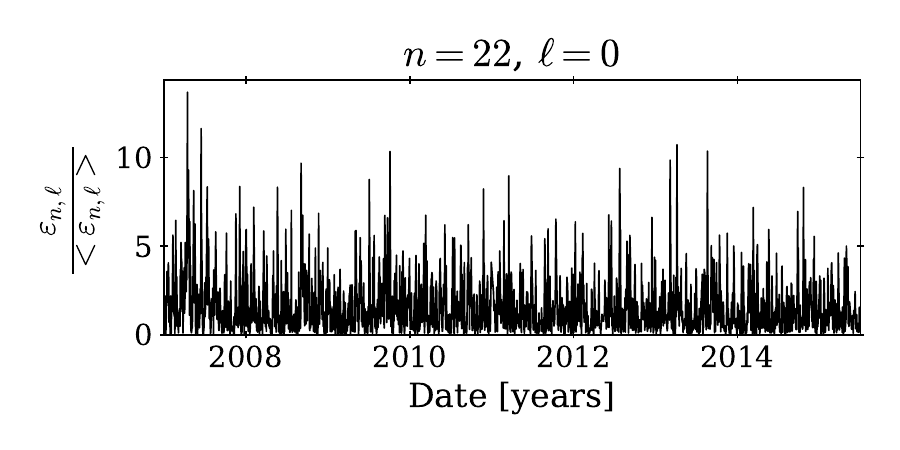}
   \caption{Normalised reconstructed energy time series for $n=22$, $\ell=0$, with VIRGO/SPMs data. For better visibility, we restricted the x-axis between 2007 and 2015 but the series were reconstructed from January 23$^{\mathrm{rd}}$, 1996 to February 17$^{\mathrm{th}}$, 2016.}
              \label{fig:energy_22-0}
    \end{figure}

All $n=14$ to $n=25$ and $\ell=0$ to $\ell=2$ modes were reconstructed using a window $\Delta\omega=8\,\mu$Hz, as recommended by F+98 and illustrated in Fig.~\ref{fig:modes_identification}. The reconstructed time series, with a time resolution of $\delta t\sim1.45$,days, allows studying short-term variations for each mode, in contrary to other classical methods. The window is centred on the mode frequencies listed in \citet{lazrek_1997}, originally derived from early GOLF data via peakbagging, and also valid for VIRGO/SPMs observations \citep{toutain_1997}. Even though the frequencies of some $\ell=3$ to $\ell=5$ modes were constrained by \citet{lazrek_1997}, we chose not to include them in this study due to their low signal-to-noise ratio. Moreover, since their frequencies are not within $\Delta\omega$ of the $\ell=0$ to $\ell=2$ modes, they do not impact our reconstruction. For comparison between modes, each time series was normalised by its mean. Figure~\ref{fig:energy_22-0} illustrates the normalised energy for $n=22$, $\ell=0$ during Solar Cycle 24 and as expected, the excitation looks stochastic during this epoch.

\subsection{Statistical tests on complete energy time series}
\label{subsec:KS_complete}

\begin{table*}
    \centering
    \caption{Starting and ending dates of the gaps larger than one day in the VIRGO/SPMs data.}
    \begin{tabular}{c|c|c}
        Start & End & Length \\
        \begin{tabular}{cc}
        Date & Hours \\
        \hline
        1996-09-09 & 15:47:34 \\
        1998-03-25 & 23:59:33 \\
        1998-06-24 & 22:16:33 \\
        1998-12-20 & 23:58:33 \\
        1999-02-14 & 12:20:32 \\
        1999-11-28 & 11:37:32 \\
        1999-12-01 & 18:35:32 \\
        2000-01-06 & 23:59:32 \\
        2000-11-28 & 23:59:32 \\
        2002-02-05 & 01:47:32 \\
        2004-04-22 & 05:47:32 \\
        2006-08-10 & 05:53:31 \\
        2011-01-02 & 03:27:30 \\
        2012-05-03 & 22:22:30 \\
        2012-05-08 & 20:23:30 \\
        2015-05-17 & 14:07:29 \\
        2017-02-13 & 23:58:27 \\
        2019-02-07 & 06:52:27 \\
        \end{tabular} &
        \begin{tabular}{cc}
        Date & Hours \\
        \hline
        1996-09-13 & 00:47:34 \\
        1998-03-27 & 00:17:33 \\
        1998-10-07 & 01:32:33 \\
        1999-02-04 & 02:50:32 \\
        1999-02-18 & 11:30:32 \\
        1999-11-29 & 20:30:32 \\
        1999-12-02 & 20:59:32 \\
        2000-01-07 & 23:59:32 \\
        2000-11-30 & 00:16:32 \\
        2002-02-08 & 01:32:32 \\
        2004-04-27 & 00:05:32 \\
        2006-08-11 & 14:52:31 \\
        2011-01-07 & 00:33:30 \\
        2012-05-04 & 23:59:30 \\
        2012-05-11 & 00:21:30 \\
        2015-05-23 & 00:34:29 \\
        2017-02-14 & 23:59:27 \\
        2019-02-12 & 00:05:27 \\
        \end{tabular} &
        \begin{tabular}{c} 
        [days] \\
        \hline
        3.38 \\   
        1.01 \\   
        104.14 \\   
        45.12 \\   
        3.97 \\   
        1.37 \\   
        1.10 \\   
        1.00 \\   
        1.01 \\   
        2.99 \\   
        4.76 \\   
        1.38 \\   
        4.88 \\   
        1.07 \\   
        2.17 \\   
        5.44 \\   
        1.00 \\   
        4.72 \\   
        \end{tabular}
        \\
        \hline
    \end{tabular}
    \label{tab:>1day_gaps}
\end{table*}

Under the hypothesis of stochastically excited modes by convection, and neglecting the mode asymmetry, each resulting time series should be distributed according to a theoretical exponential distribution $f_{\mathrm{th}}$ defined by its mean $\lambda$ for positive values of $x$:
\begin{equation}
    f_{\mathrm{th}}(x) = \frac{1}{\lambda} \exp{\left(-\dfrac{x}{\lambda}\right)}\,,
    \label{eq:exponential_distribution}
\end{equation}
and this is confirmed by Eq.~\eqref{eq:energy} assuming that the distribution of oscillatory velocities is Gaussian.

\begin{figure}
   \centering
   \includegraphics[width = 0.5\textwidth,trim={0.5cm 0 0 0},clip]{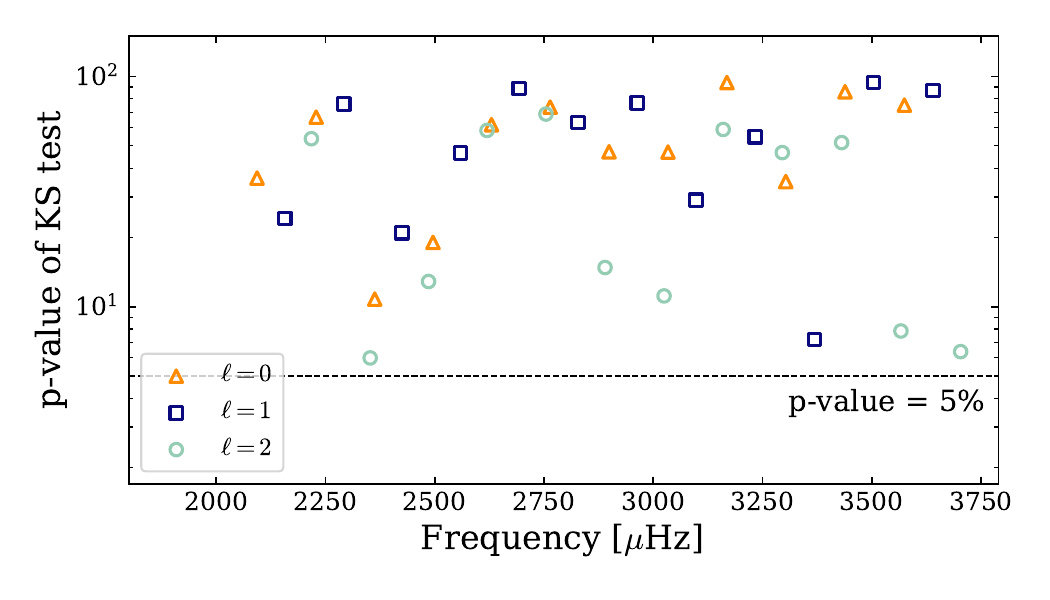}
   \caption{Resulting p-values of the Kolmogorov-Smirnov tests comparing reconstructed time series distributions with exponential distributions. 
   Orange triangles stand for $\ell=0$ modes, dark blue squares for $\ell=1$ modes and light blue circles for $\ell=2$ modes. The threshold of rejection of 5\,\% is drawn in dashed black.}
        \label{fig:KS_expon_fullts}
    \end{figure}

We performed Kolmogorov-Smirnov (KS) tests to compare for each mode, the agreement between the reconstructed time series and $f_{\mathrm{th}}$. This statistical test compares a sample to a theoretical distribution \citep{massey_1951} by evaluating several statistics including the maximum distance between both cumulative distributions and the p-value. The null hypothesis $H_0$ is that the observed distribution follows the theoretical one. In this context, the p-value estimates the probability of being wrong in rejecting $H_0$. As is commonly done, we set a 5\,\% threshold for the p-value in order to choose whether or not to reject $H_0$. 
All statistical tests were conducted excluding values associated with gaps longer than one day in the original time series. Additionally, a data point at the edges before and after these gaps were removed to prevent potential edge effects during the reconstruction. The dates of these gaps are listed in Table~\ref{tab:>1day_gaps}.
The values for $\lambda$ were estimated via Maximum Likelihood Estimation (MLE) from the 6858 points of each dataset, by minimizing the negative log-likelihood function defined as:
\begin{equation}
    \log L(\lambda) = -n\log\lambda -\frac{1}{\lambda}\sum_{i=1}^{n}x_i\,,
\end{equation}
where $n$ is the total number of observations and $x_i$ are the observed data points.
Resulting p-values of the KS tests are shown in Fig.~\ref{fig:KS_expon_fullts} as well as the rejection threshold of 5\,\%. No points are below this threshold: all distributions correspond to modes which are not incompatible with exponential distributions according to the test. 

\subsection{Statistical tests on activity-dependent subsamples}
\label{subsec:KS_activity}

\begin{figure}
   \centering
   \includegraphics[width = 0.5\textwidth,trim={0.5cm 0 0 0},clip]{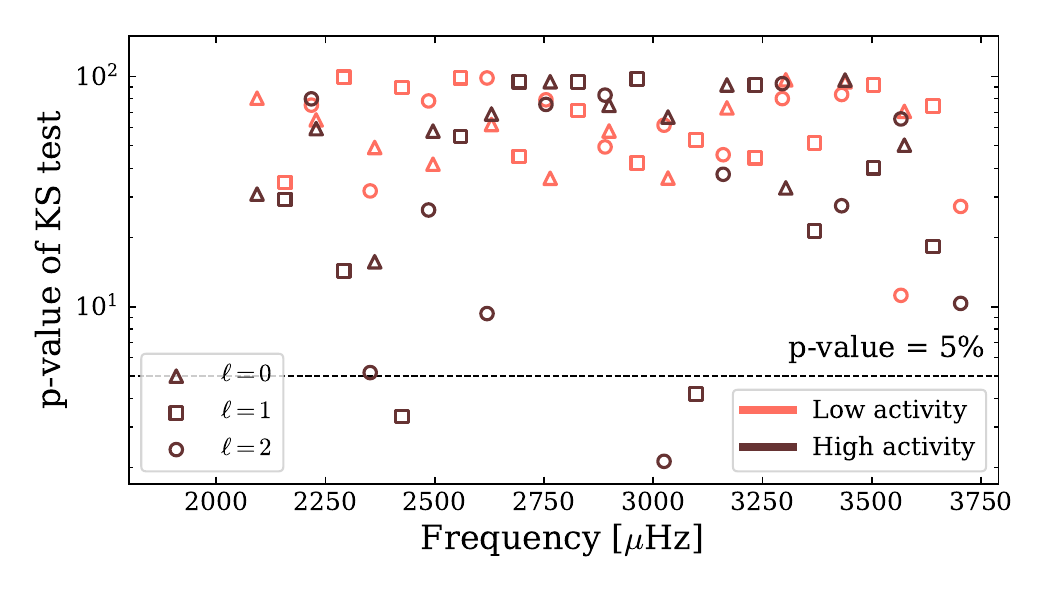}
   \caption{Resulting p-values of the Kolmogorov-Smirnov tests comparing reconstructed time series distributions with exponential distributions, for measurements taken during low and high magnetic activity periods whose dates are listed in Table~\ref{tab:cycle_extrema}. Triangles stand for $\ell=0$ modes, Squares for $\ell=1$ modes and Circles for $\ell=2$ modes. Brown and pink makers stand for tests on values taken during high (resp. low) magnetic activity. The threshold of rejection of 5\,\% is drawn in dashed black.}
        \label{fig:KS_expon_extrema}
    \end{figure}

Given that the mode parameters are affected by solar activity, we repeated the statistical analysis of the mode energy using sub-samples corresponding to low and high levels of magnetic activity.
Among the various activity proxies discussed in \citet{jain_2009}, the solar radio flux index $F_{10.7}$ \citep{tapping_2013}, sensitive to the upper chromosphere, shows the strongest correlation with p-mode frequency shifts over full solar cycles. Based on these findings, $F_{10.7}$ is selected as our reference proxy to define epochs of low and high activity (see Appendix~\ref{app:F10.7}). Table~\ref{tab:cycle_extrema} summarises the corresponding dates for solar minima and maxima.

\begin{table}
  \caption{Thresholds used to define starting and ending dates of each cycle extremum for Cycles 23, 24 and 25. Dates are also listed in the two right columns.}
     \label{tab:cycle_extrema}
     \centering
     \begin{tabular}{lcc}
        \hline
        \noalign{\smallskip}
        Cycle extremum & Start & End \\
        \noalign{\smallskip}
        \hline
        \noalign{\smallskip}
        Minimum 23 & 1995-08-16 & 1997-06-19  \\
        Maximum 23 & 2000-02-03 & 2002-10-08  \\
        Minimum 24 & 2007-02-24 & 2009-12-05  \\
        Maximum 24 & 2011-09-23 & 2015-02-11 \\
        Minimum 25 & 2016-12-16 & 2021-04-04  \\
        Maximum 25 & 2023-01-13 & 2024-07-16\tablefootmark{(a)}  \\
        \noalign{\smallskip}
        \hline
     \end{tabular}
\tablefoot{
    \tablefoottext{a}{Maximum of Cycle 25 is the current phase: we stopped the analysis on July 16th, 2024.}
    }
\end{table}

Each reconstructed time series was divided into two subsets: one for low activity (1735 points) and one for high activity (2125 points). KS tests were then performed separately to each and figure~\ref{fig:KS_expon_extrema} shows the resulting p-values. Three p-values fall below the 5\,\% threshold, all corresponding to high-activity intervals. Given that 72 tests were conducted, observing a few rejections is statistically expected. However, the fact that all rejections occur during high activity periods suggests a possible, though subtle, deviation from the stochastic model during these phases.

\subsection{Statistical study of non-stochasticity in the modes excitation}
\label{subsec:gamma_dist}

\begin{figure}
    \centering
    \includegraphics[width = 0.5\textwidth]{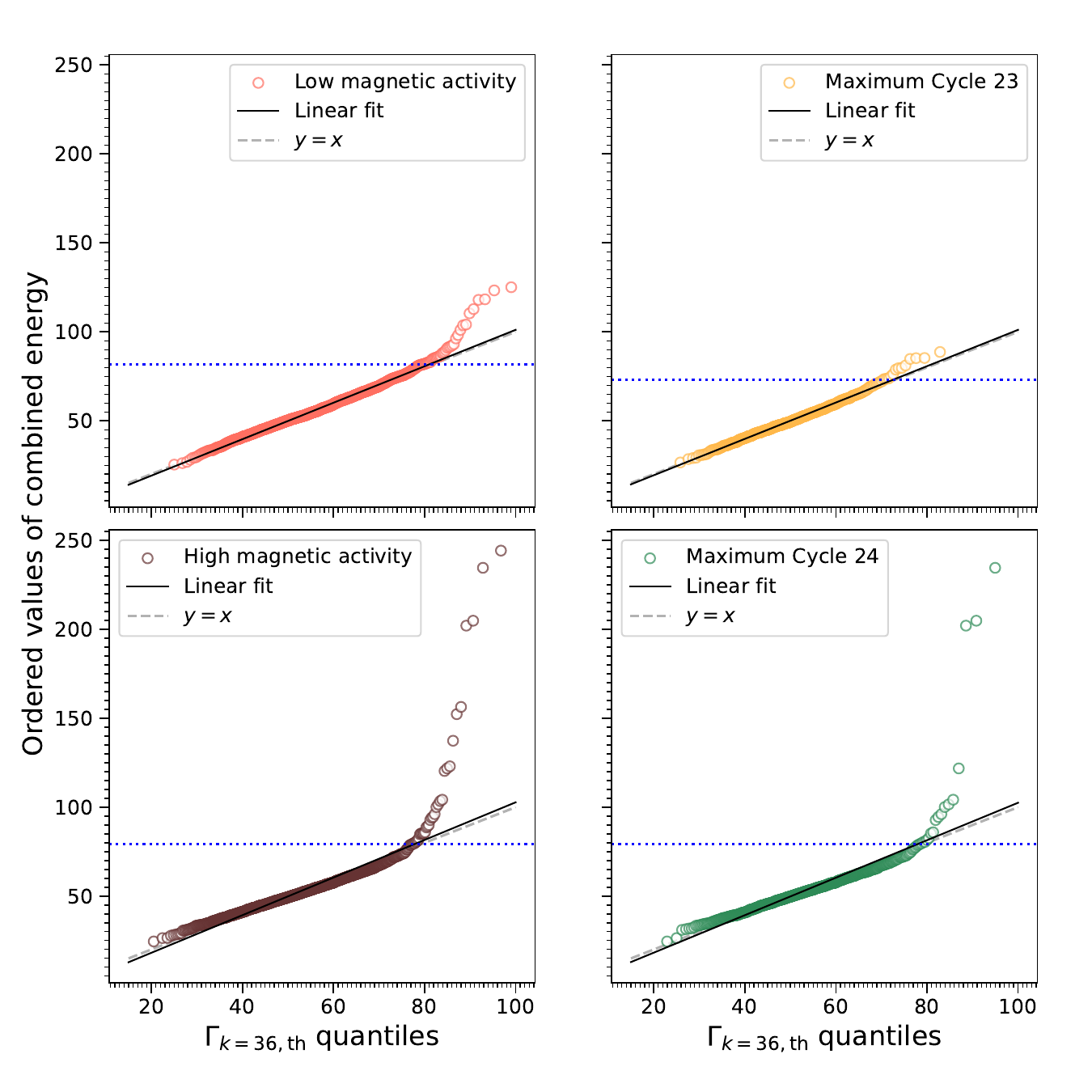}
    \caption{Q-Q plot comparing the combined energy time series with $\varGamma_{k=36, \mathrm{th}}$ distributions. Panels correspond to the 4 selected sub-samples: low magnetic activity (in pink, top left), high magnetic activity (in brown, bottom left), maximum of Cycle 23 (in yellow, top right) and maximum of Cycle 24 (in green, bottom right). The general trend of the Q-Q plot was fitted with a least square regression to the data and is shown as a black line on all panels. The grey dashed line shows the line $y=x$, and the blue dotted line shows the value of the quantile at 98\,\% for each period.}
        \label{fig:QQ-plot}
\end{figure}

\begin{table}
    \centering
    \caption{Resulting p-values of the KS tests performed to compare the combined energy time series with the $\varGamma_{k, \mathrm{th}}$ distributions, considering 4 time periods. $\alpha$ and $\beta$ parameters of the Gamma-distribution determined by Maximum Likelihood Estimation for each sub-sample are also listed.}
    \begin{tabular}{lccc}
        Period & p-value (\,\%) & $\alpha$ & $\beta$ \\
        \hline
        Low activity & 9.61 & -9.21\,\footnotesize{$\pm0.98$} & 1.80\,\footnotesize{$\pm0.03$}\\
        High activity & $7.7\,10^{-5}$ & -15.75\,\footnotesize{$\pm1.18$} & 1.89\,\footnotesize{$\pm0.03$} \\
        Maximum 23 & 64.41 & -5.42\,\footnotesize{$\pm1.49$} & 1.54\,\footnotesize{$\pm0.04$} \\
        Maximum 24 & 0.14 & -15.15\,\footnotesize{$\pm1.70$} & 1.90\,\footnotesize{$\pm0.05$} \\
        \hline
    \end{tabular}
    \label{tab:p-values_KStests_gammadist}
\end{table}

To explore potential deviations from purely stochastic excitation, we combined all 36 normalised mode energy time series into a single sequence. Under the hypothesis of independent, exponentially distributed signals, the resulting distribution should follow a Gamma-distribution of order $k$, corresponding to the number of combined modes. Further KS tests were performed for four subsets of the sequence: low and high activity, both already considered in Sect.~\ref{subsec:KS_activity}, and the maxima of Cycles 23 and 24, separately.
We considered the parametrised Gamma-distribution noted $\varGamma_{k, \mathrm{th}}$, defined by:
\begin{equation}
    \varGamma_{k, \mathrm{th}}(x, \alpha, \beta) = \frac{(x-\alpha)^{k-1}e^{-(x-\alpha)/\beta}}{\beta^{k}\,\Gamma(x)}\,,
    \label{eq:gamma-distribution}
\end{equation}
where $\Gamma(x)$ denotes the Euler Gamma function, and $\alpha$ and $\beta$ are respectively the location and the scale parameters. Fixing $k=36$, $\alpha$ and $\beta$ parameters were fitted via MLE by minimizing:
\begin{multline}
    \log L(k=36, \alpha, \beta) = n[k\log\beta-\log\Gamma(k)] \\
    - \sum_{i=1}^{n}(k-1)\log(x_i-\alpha)-\frac{1}{\beta}\sum_{i=1}^{n}(x_i-\alpha)\,.
\end{multline}
Fitted values for $\alpha$ and $\beta$ are listed in Table~\ref{tab:p-values_KStests_gammadist} along with the resulting p-values of the KS tests.

\begin{figure}
    \centering
    \includegraphics[width = 0.5\textwidth,trim={0 0 0 0},clip]{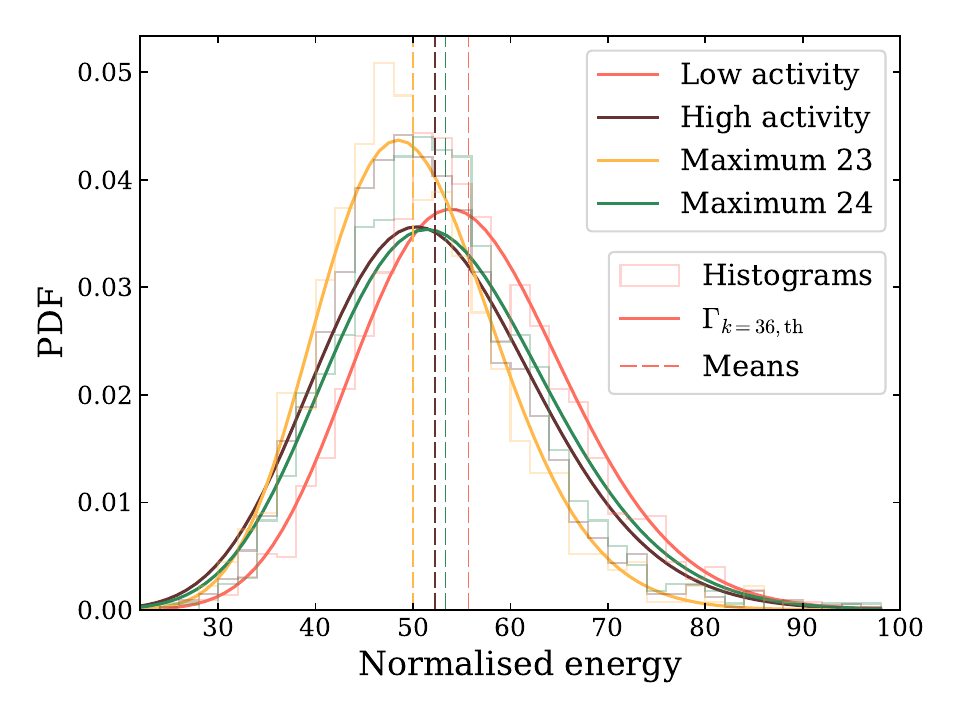}
    \caption{Distributions of normalised combined energies. For each cycle period, we show the histogram, the best $\varGamma_{k=36,\mathrm{th}}$ distribution in continuous line and its mean in dashed line. Red data correspond to low magnetic activity, the brown ones to high activity, the yellow ones to the maximum of Cycle 23 and the green ones to the maximum of Cycle 24.}
        \label{fig:gamma_dist}
\end{figure}

Results statistically consistent with H$_0$ are obtained only for the low-activity phase and the maximum of Cycle 23 subsets. The p-value for the Cycle 23 is indeed close to 50\,\%, while for Cycle 24 it is below the 5\,\% threshold. 
The constrained estimates of the parameters $\alpha$ and $\beta$ indicate a clear divergence in mode energy distributions between low- and high-activity subsets, as well as between the maxima of Cycles 23 and 24.
This discrepancy is also evident in the quantile-quantile (Q-Q) plots presented in Fig.~\ref{fig:QQ-plot}, where data points above the 98\,\% quantile systematically deviate from the theoretical distributions, especially in subsets with lower KS test p-values.
The deviations from the $y=x$ reference line are more pronounced for high-activity phases compared to low-activity ones, and for the maximum of Cycle 24 relative to the maximum of Cycle 23.  
These findings suggest that the statistical properties of mode energy during periods of high- magnetic activity are significantly influenced by the distinct behaviour of Cycle 24, and that there is a difference between the maximum of Cycle 23 and the one of Cycle 24, corroborating previous results reported by e.g. \citet{garcia_2024}.
Figure~\ref{fig:gamma_dist} shows the corresponding Gamma-distributions. As expected from the anti-correlation between mode energy and magnetic activity, the mean combined energy is higher during low-activity periods than during high-activity ones. The mean normalised energy is also lower during the maximum of Cycle 23 than that of Cycle 24, again highlighting differences between the two cycles. Moreover, variance likewise varies with activity: it is smaller during the Cycle 23 maximum compared to low activity, but larger during the Cycle 24 maximum.

To quantify the high-energy excess seen in Fig.~\ref{fig:QQ-plot}, we analysed short-timescale fluctuations in energy by dividing the combined time series by its 1-year moving mean. This residual energy time series is further noted $\mathcal{S}_{\mathrm{phot}}$. The high-energy excesses are defined as the local maxima in $\mathcal{S}_{\mathrm{phot}}$ that have energies exceeding the threshold $x_\mathrm{phot}=24.75$, that corresponds to the 98\,\% quantile of $\mathcal{S}_{\mathrm{phot}}$.
The best-fit Gamma distribution for $\mathcal{S}_{\mathrm{phot}}$ yields $\alpha_\mathrm{phot} = -58.05,\pm0.49$ and $\beta_\mathrm{phot} = 1.59,\pm0.02$, with a p-value of $1.63 \times 10^{-3},\,\%$, indicating a strong deviation from the theoretical model. After removing peaks above the threshold, the p-value rises to 30.15\,\%, confirming that high-energy events are the source of the deviation.
The expected probability of observing events above the given threshold $x$ can be estimated by applying the binomial probability formula:
\begin{equation}
    P(r;p,n) = p^r (1-p)^{n-r} \frac{n!}{r!(n-r)!}\,,
    \label{eq:binomial_probability}
\end{equation}
with $r = 108$, $n = 6612$ and $p = \varGamma_{k=36,\mathrm{th}}(x_\mathrm{phot}, \alpha_\mathrm{phot}, \beta_\mathrm{phot})$. We find $P \ll 10^{-3}$: the observed number of high-energy events is significantly higher than expected from a stochastic model. Our results echo previous studies by \citep{chaplin_1997, chang_1998} who found a power excess in the modes when working with a total of $\sim763.8$\,days of the Birmingham Solar-Oscillations Network (BiSON) data, collected in short segments of 11.4 h.
We will discuss further the found excess of energy in Sect.~\ref{sec:solutions}.

\begin{figure}
    \centering
    \includegraphics[width=0.48\textwidth,trim={2 0 0 0},clip]{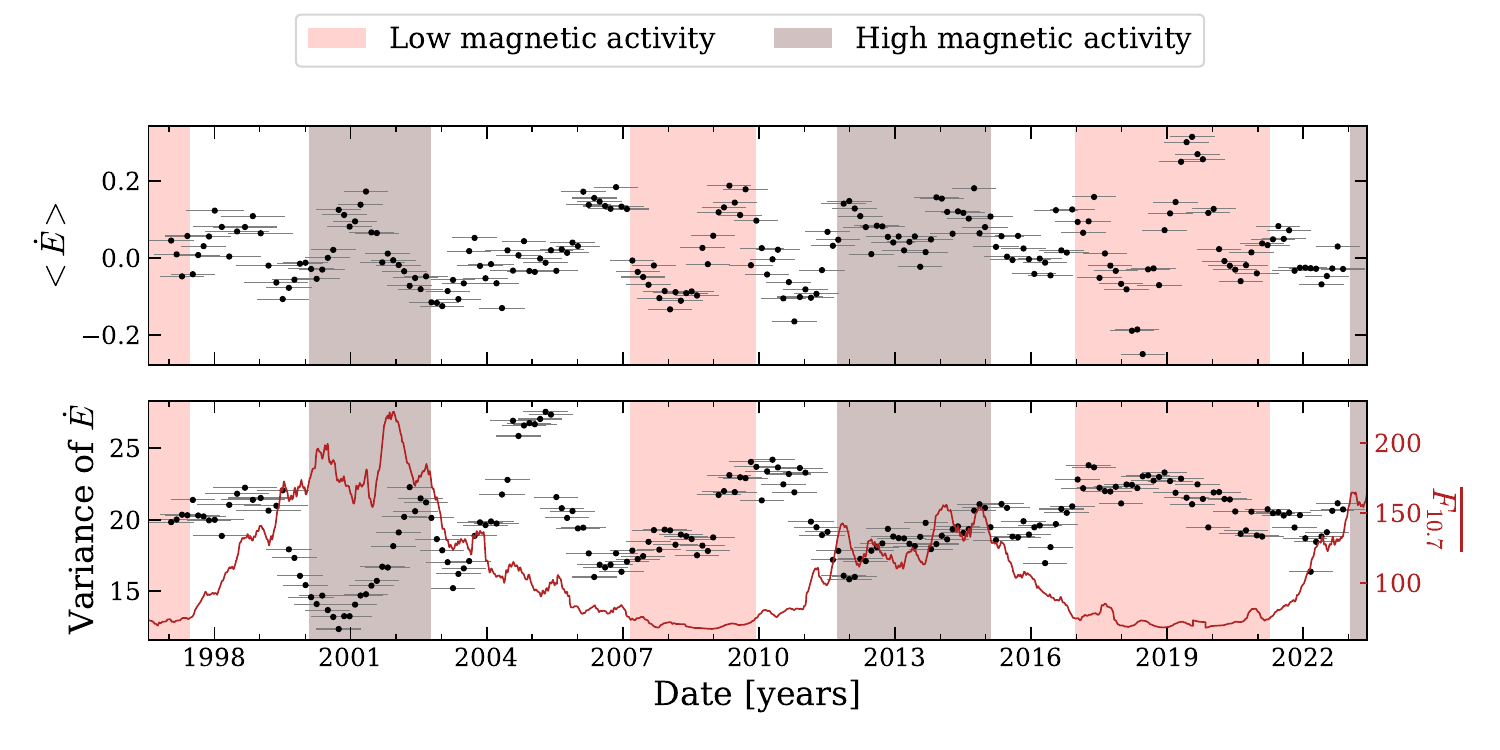}
    \caption{Mean (top panel) and variance (bottom panel) of 1-year subseries slided every eighth of a year of the energy supply rate computed from the combined time series of the modes. These quantities are plotted against the central date of the considered subseries and the x-axis error bars show the range of dates taken for each subseries. Periods of low (pink) and high (brown) magnetic activity were reported from Table~\ref{tab:cycle_extrema} and $\overline{F_{10.7}}$ is reported in red, in the bottom panel, for comparisons.}
    \label{fig:dEdt_variations}
\end{figure}

Now focusing on the energy below the threshold $x_\mathrm{phot}$ (statistically consistent with a stochastic excitation), we can compute the energy supply rate \citep[$\dot{E}=\mathrm{d}E/\mathrm{d}t$, e.g.][]{jimenez-reyes_2003}. 
The central finite difference formula giving a second-order approximation of the elements $i$ of the first derivative of a function $f(x)$, can be adapted to discrete functions with a non-constant step size with:
\begin{multline}
        f'(x_i)\approx \frac{h_{i-1}^2 f(x_{i+1}) + (h_i^2-h_{i+1}^2) f(x_i)-h_i^2f(x_{i-1})}{h_{i-1}h_i(h_i+h_{i-1})} \\+ \mathcal{O} \left( \frac{h_i h_{i-1}^2+h_i^2h_{i-1} }{h_i+h_{i-1}} \right)\,,
\end{multline}
where $f(x)$ is supposed three times differentiable, $h_i = x_{i+1}-x_i$ and $h_{i-1} = x_{i}-x_{i-1}$ \citep{quarteroni_2007}.
The temporal mean and variance of $\dot{E}$ were computed over 1-year sub-series, shifted every eighth of a year.
The results are illustrated in Fig.~\ref{fig:dEdt_variations}. 
The variation in the mean energy supply rate follows a quasi-biennial oscillation (QBO) pattern, consistent with previous findings in mode frequency shifts and magnetic proxies \citep[][and references therein]{bazilevskaya_2014}.
The variance is anti-correlated with solar activity and shows stronger fluctuations during Cycle 24 than Cycle 23, in agreement with similar observations from high-degree mode analyses using GONG data \citep{kiefer_2021}.

\section{Possible explanations for energy excess}
\label{sec:solutions}

The time series of the combined energy of the modes reveals an excess of energy in the modes compared to the expected level under the hypothesis of stochastic excitation by turbulent convection. 
To better understand this discrepancy, several potential sources were investigated.
Here, we analyse the temporal correlation of the 108 high-energy points with a set of indicators: the occurrence of energetic flares and CMEs, the reconstructed mode energy from GOLF observations, and the potential influence of either photon noise in the VIRGO/SPM instruments or convective processes.
Analysing the sensitivity depth of the modes, the possible location of their non-stochastic excitation is also discussed. After detailing the data used in Sect.~\ref{subsec:data_comparisons} for which the notations are summarised in Table~\ref{tab:summary_notations}, we explore the potential sources of high-energy peaks found in $\mathcal{S}_{\mathrm{phot}}$ in Sect.~\ref{subsec:analyse_comparison}.

\subsection{Complementary indicators}
\label{subsec:data_comparisons}

\begin{table*}
    \centering
    \caption{Summary of the notations introduced for the discussed time series. From left to right, the columns provide the notation, the instruments from which the data were obtained, the frequencies of the considered modes, the number of modes, the total number of peaks counted above the quantile at 98\,\% of each time series and the reference sections detailing the construction of each of these time series.}
    \begin{tabular}{lccccc}
        \multirow{ 2}{*}{Name} & \multirow{ 2}{*}{Instrument} & Frequencies & Number & Number & \multirow{ 2}{*}{Section} \\
        & & [$\mu$Hz] & of modes & of peaks & \\
        \hline
        $\mathcal{S}_{\mathrm{phot}}$ & VIRGO/SPMs & 2090-3710 & 36 modes & 108 & Sect.~\ref{subsec:gamma_dist}\\
        $\sigma_{\mathrm{phot}}$ & VIRGO/SPMs & 6590-8210 & 36 windows of photon noise & 92 & Sect.~\ref{subsubsec:photon_noise} \\
        $\mathcal{C}_{\mathrm{phot}}$ & VIRGO/SPMs & 90-1710 & 36 windows of convective contribution & 103 & Sect.~\ref{subsubsec:photon_noise} \\
        $\mathcal{L}_{\mathrm{phot}}$ & VIRGO/SPMs & 2090-2620 & 12 modes & 121 & Sect.~\ref{subsubsec:fb}\\
        $\mathcal{M}_{\mathrm{phot}}$ & VIRGO/SPMs & 2620-3160 & 12 modes & 115 & Sect.~\ref{subsubsec:fb}\\
        $\mathcal{H}_{\mathrm{phot}}$ & VIRGO/SPMs & 3160-3710 & 12 modes & 123 & Sect.~\ref{subsubsec:fb}\\
        $\mathcal{S}_{\mathrm{vel}}$ & GOLF & 2090-3710 & 36 modes & 76 & Sect.~\ref{subsubsec:golf}\\
        \hline
    \end{tabular}
    \label{tab:summary_notations}
\end{table*}

\subsubsection{Contribution from other frequency scales}
\label{subsubsec:photon_noise}

Since the reconstruction is sensitive to the photon noise level, we applied the same reconstruction procedure, using the mode pattern shifted to higher frequencies, i.e. adding 4500\,$\mu$Hz to each mode’s central frequency. This yielded 36 windows of 8\,$\mu$Hz between 6590 and 8210\,$\mu$Hz. Each time series was normalised by the mean energy of the corresponding mode in $\mathcal{S}_{\mathrm{phot}}$, and the resulting 36 series were combined into a new time series, $\sigma_{\mathrm{phot}}$, in which 92 points exceed the 98\,\% quantile threshold.

An analogous reconstruction was performed with a downward shift of 2000\,$\mu$Hz, yielding an energy time series dominated by convective contributions, denoted $\mathcal{C}_{\mathrm{phot}}$, in which 103 points lie above the threshold.

\subsubsection{Flares and CMEs}
\label{subsubsec:flares_CMEs}

Flares are classified based on their peak X-ray flux in the 1–8\,$\mathring{\text{A}}$ range. X-class flares (with a flux greater than $10^{-4}\,\mathrm{W}\,\mathrm{m}^{-2}$) correspond to energy releases larger than $10^{32}$\,ergs, and were extracted from GOES satellite data\footnote{\url{ftp://ftp.ngdc.noaa.gov/STP/space-weather/solar-data/solar-features/solar-flares/x-rays/goes/}} \citep[Geostationary Operational Environmental Satellite,][]{mallette_1982}. Samely, the top 5\,\% most energetic CMEs (with an energy greater than $1.3 \times 10^{31}$\,ergs) were selected from SoHO/LASCO data\footnote{\url{https://cdaw.gsfc.nasa.gov/CME_list/}} \citep[Large Angle and Spectrometric COronagraph experiment,][]{brueckner_1995}. Over the VIRGO/SPMs operational period, 1271 CMEs and 195 X-class flares were recorded.

\subsubsection{Modes energy time series from GOLF data}
\label{subsubsec:golf}

\begin{figure}
    \centering
   \includegraphics[width = 0.5\textwidth,trim={0 0 0 0},clip]{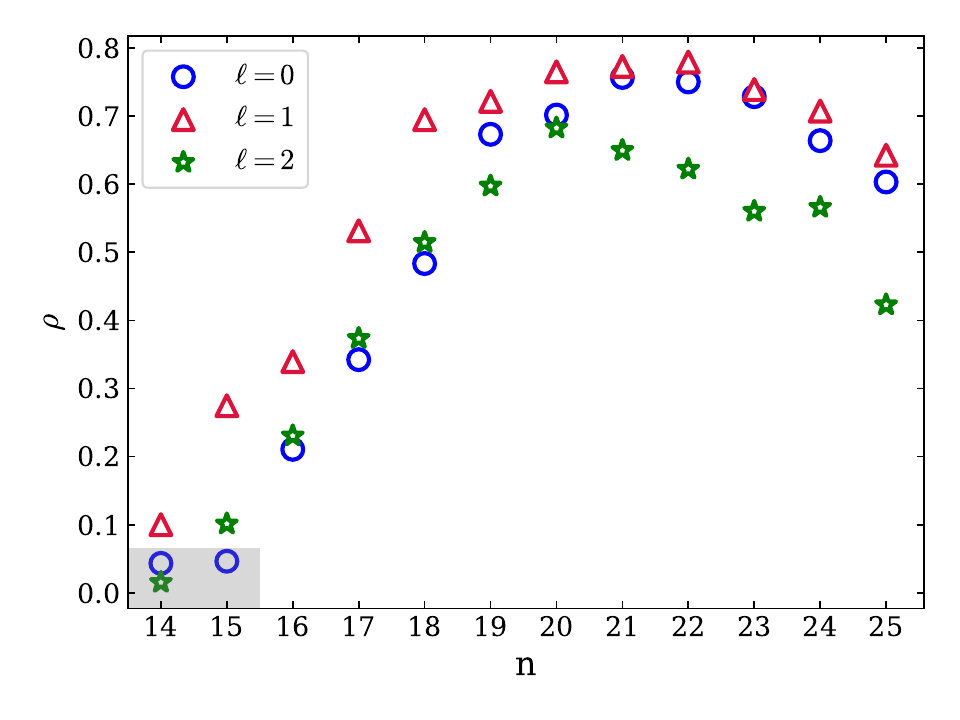}
   \caption{Spearman’s rank correlation $\rho$ between $\mathcal{S}_{\mathrm{phot}}$ and $\mathcal{S}_{\mathrm{vel}}$ data, versus radial order $n$ for all considered modes. Blue circles stands for $\ell=0$ modes, red triangles for $\ell=1$ and green stars for $\ell=2$. Markers in the grey area correspond to a p-value larger than $10^{-3}\,\%$.}
        \label{fig:spearman_correlation}
\end{figure}

Energy time series for the same modes were reconstructed from GOLF data starting on November 18\textsuperscript{th}, 2002, when the instrument was switched back to its blue-wing configuration. 
These series were normalised by their 365-day moving mean to correct for the annual modulation in photon noise \citep[see Appendix~\ref{app:calibration_VIRGO} and][]{garcia_2005}. 
Spearman's rank correlations were then computed between $\mathcal{S}_{\mathrm{phot}}$ and the GOLF time series, as shown in Fig.~\ref{fig:spearman_correlation}. 
With a rejection threshold at 5\,\%, all modes except $n=14$, $\ell=2$ were significantly correlated. The correlation strength increases with the power of the mode, and therefore with its signal-to-noise ratio. This discrepancy between the datasets is consistent as the convective background is higher in VIRGO/SPMS than in GOLF, resulting in different amplitudes of the power spectral density (PSD), particularly for the lower frequency modes. The combined time series from GOLF, denoted $\mathcal{S}_{\mathrm{vel}}$, exhibits 76 peaks above the 98\,\% quantile.

\subsubsection{Combined time series in three frequency bands}
\label{subsubsec:fb}

Because acoustic modes are sensitive to different regions of the star depending on their frequency, we grouped the reconstructed modes into three frequency bands: 2090–2620\,$\mu$Hz, 2620–3160\,$\mu$Hz, and 3160–3710\,$\mu$Hz, each containing 12 modes, and respectively denoted $\mathcal{L}_{\mathrm{phot}}$, $\mathcal{M}_{\mathrm{phot}}$, and $\mathcal{H}_{\mathrm{phot}}$ when filtered by their one-year moving mean. 
The computation of the mean sound-speed kernels $<\mathcal{K}>$, which indicate the depth of maximum sensitivity for each band, is described in Appendix~\ref{app:kernels}.
The estimated depths of maximum sensitivity are 1895\,km for $\mathcal{H}_{\mathrm{phot}}$, 1014\,km for $\mathcal{M}_{\mathrm{phot}}$, and 605\,km for $\mathcal{L}_{\mathrm{phot}}$ beneath the surface.

\begin{table*}
    \centering
    \caption{Resulting p-values of the KS tests performed to compare combined energy time series in the three frequency bands 2090-2620\,$\mu$Hz, 2620-3160\,$\mu$Hz, and 3160-3710\,$\mu$Hz, with the best $\varGamma_{k=12, \mathrm{th}}$ distributions whose parameters $\alpha$ and $\beta$ were estimated from Maximum Likelihood Estimation.}
    \begin{tabular}{l|c|ccc}
        \multirow{3}{*}{\textbf{Periods}} & \multirow{3}{*}{\textbf{Parameters}} & \multicolumn{3}{c}{\textbf{Band} $[\mu$Hz$]$} \\[0.2cm]
         &  & \textbf{2090-2620} & \textbf{2620-3160} & \textbf{3160-3710} \\[0.2cm]
        \hline\hline
        &\\[-0.2cm]
        \multirow{3}{1.5cm}{\textbf{Low activity}} & p-value & 85.3\,\% & 98.6\,\% & 44.5\,\% \\
         & $\alpha$ & -0.97\,\footnotesize{$\pm0.33$} & -1.57\,\footnotesize{$\pm0.35$} & -0.82\,\footnotesize{$\pm0.33$} \\
         & $\beta$ & 1.59\,\footnotesize{$\pm0.03$} & 1.70\,\footnotesize{$\pm0.03$} & 1.62\,\footnotesize{$\pm0.03$} \\[0.2cm]
        \hline
        &\\[-0.2cm]
        \multirow{3}{1.5cm}{\textbf{High activity}} & p-value & 0.15\,\% & 31.3\,\% & 53.4\,\% \\
         & $\alpha$ & -2.83\,\footnotesize{$\pm0.34$} & -1.08\,\footnotesize{$\pm0.27$} & -1.29\,\footnotesize{$\pm0.29$} \\
         & $\beta$ & 1.75\,\footnotesize{$\pm0.03$} & 1.49\,\footnotesize{$\pm0.03$} & 1.54\,\footnotesize{$\pm0.03$} \\[0.2cm]
        \hline
        &\\[-0.2cm]
        \multirow{3}{1.5cm}{\textbf{Maximum Cycle 23}} & p-value & 92.1\,\% & 47.0\,\% & 99.5\,\% \\
         & $\alpha$ & -0.76\,\footnotesize{$\pm0.45$} & -0.47\,\footnotesize{$\pm0.40$} & -0.24\,\footnotesize{$\pm0.41$} \\
         & $\beta$ & 1.51\,\footnotesize{$\pm0.04$} & 1.37\,\footnotesize{$\pm0.04$} & 1.41\,\footnotesize{$\pm0.04$} \\[0.2cm]
        \hline
        &\\[-0.2cm]
        \multirow{3}{1.5cm}{\textbf{Maximum Cycle 24}} & p-value & 8.8\,\% & 25.8\,\% & 52.8\,\% \\
         & $\alpha$ & -2.58\,\footnotesize{$\pm0.49$} & -1.06\,\footnotesize{$\pm0.41$} & -1.81\,\footnotesize{$\pm0.44$} \\
         & $\beta$ & 1.74\,\footnotesize{$\pm0.04$} & 1.52\,\footnotesize{$\pm0.04$} & 1.62\,\footnotesize{$\pm0.04$} \\[0.2cm]
         \hline
    \end{tabular}
    \label{tab:p-values_KStests_gammadist_per_FB}
\end{table*}

\begin{figure}
    \centering
    \includegraphics[width=0.5\textwidth,trim={0.9cm 0.9cm 0 0},clip]{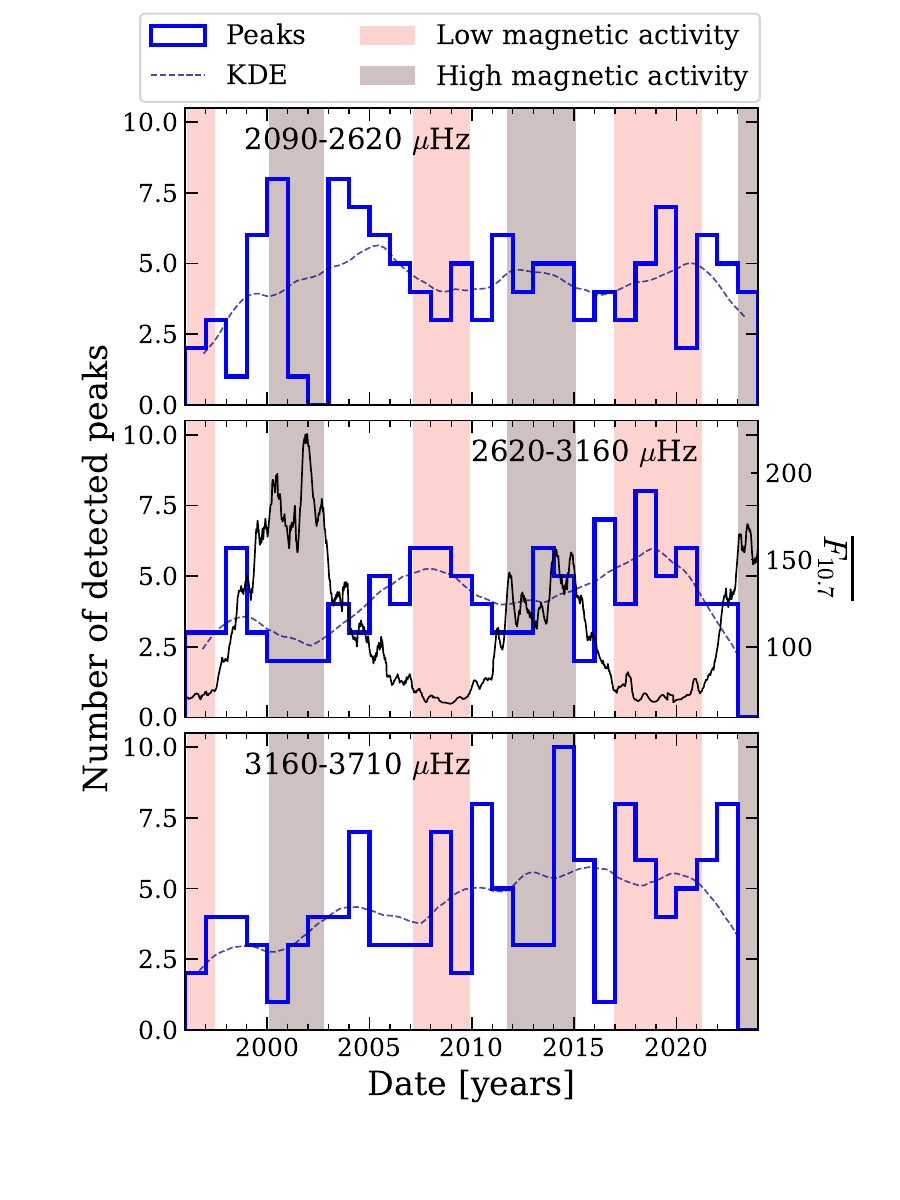}
    \caption{Distribution of the detected peaks over time for the three frequency bands, from top to bottom: 2090-2620\,$\mu$Hz, 2620-3160\,$\mu$Hz, and 3160-3710\,$\mu$Hz. Each bin corresponds to one year from 1995 to 2025. Dashed curve shows the Epanechnikov Kernel Density Estimations (KDE) of the probability distributions from each histogram. Periods of low and high magnetic activity were reported from Table~\ref{tab:cycle_extrema} and $\overline{F_{10.7}}$ is reported in black, in the middle panel, for comparisons.}
    \label{fig:peaks_over_time_fb}
\end{figure}

KS tests were performed comparing each band’s time series to theoretical Gamma-distributions for the same subsets described in Sect.~\ref{subsec:gamma_dist}. Table~\ref{tab:p-values_KStests_gammadist_per_FB} summarises the resulting p-values, and the estimated $\alpha$ and $\beta$ parameters, this time by fixing $k=12$ in Eq.~\eqref{eq:gamma-distribution}. 
A similar pattern to that observed in the total time series is found across all frequency bands, with smaller p-values during the maximum of Cycle 24 than during the maximum of Cycle 23. Notably, in the low-frequency band, the p-value drops below the 5\,\% threshold during periods of high magnetic activity. In contrast, the high-frequency band shows a higher p-value during high activity than during low activity, whereas the opposite behaviour is seen in the two lower-frequency bands.
Peaks exceeding the 98\,\% quantile were identified for the three time series, yielding 121, 115, and 123 peaks in the low-, medium-, and high-frequency bands, respectively. Their temporal distribution is shown in Fig.~\ref{fig:peaks_over_time_fb}. The number of high energy peaks in the medium-frequency band notably exhibit an anti-correlation with the solar cycle, suggesting a possible excitation mechanism linked to solar activity. This anti-correlation is absent in the other two bands, a difference possibly related to their lower signal-to-noise ratios.

\subsection{Analysis and discussion}
\label{subsec:analyse_comparison}

We now examine potential sources behind the observed high-energy excess in the modes.
Key numbers in this section are illustrated in Fig.~\ref{fig:summary_comparisons}, and listed in Appendix~\ref{app:table_VIRGO_peaks}.

\begin{figure*}
    \includegraphics[width = 1\textwidth,trim={4.5cm 2cm 3.5cm 0},clip]{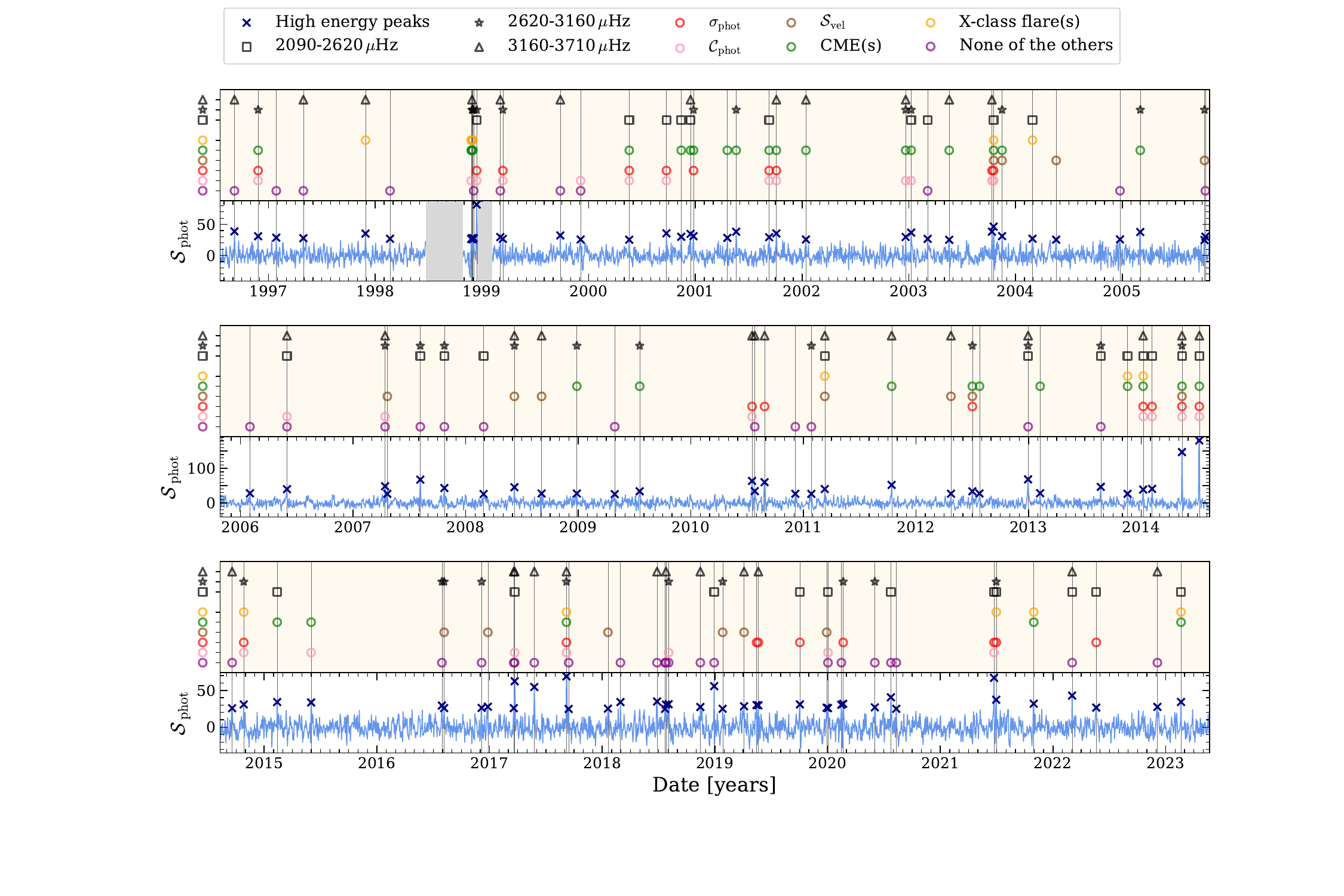}
   \caption{$\mathcal{S}_{\mathrm{phot}}$ is shown in light blue in the lower part of each panel, along with the detected high-energy peaks marked by dark blue crosses. To enhance clarity, the full 27-year time series is divided into three panels (from top to bottom). For each peak, when it corresponds to a peak in $\sigma_{\mathrm{phot}}$ (resp. in $\mathcal{C}_{\mathrm{phot}}$ or in $\mathcal{S}_{\mathrm{vel}}$), within a window of $\delta t = 1.45$ days, a red (resp. pink or brown) circle highlights it in the top light beige area of each panel. Similarly, when at least one X-class flare (or one CME) occurred within the $\delta t$-window around the peak, it is marked by an orange (or green) circle. Peaks not coinciding with any of these events are marked by violet circles. Furthermore, if a peak is detected within one of the three defined frequency bands, it is highlighted with a black marker: a square for low frequencies, a star for medium frequencies, and a triangle for high frequencies. Gaps due to the SoHO recovery mission are represented as grey shaded areas. Minor ticks represent months.}
        \label{fig:summary_comparisons}
\end{figure*}

\subsubsection{External sources affecting the measurements}

Calibration of the VIRGO/SPMs and GOLF data includes sigma-clipping to remove outliers, such as flare and CME contributions \citep{garcia_2005}. However, uncorrected external events (e.g., particle hits) would elevate photon noise across all frequencies. Of the 108 high-energy peaks in $\mathcal{S}_{\mathrm{phot}}$, 26 coincide within $\delta t = 1.45$ days with a peaks in $\sigma_{\mathrm{phot}}$.
Additionally, 34 peaks occurred simultaneously (within a window of $\delta t=1.45$\,day) with at least one CME, and 14 with at least one X-class flare. 
For example, Fig.~\ref{fig:summary_comparisons} shows that an X-class flare coincided with an energetic peak in $\mathcal{S}_{\mathrm{phot}}$ in early 2004.
However, no correlation was found between the energy associated to the event and the occurrence of a mode energy peak. 
Among the 34 CME-related peaks, 11 are also present in $\sigma_{\mathrm{phot}}$, as are 5 of the 14 flare-related peaks. In total, 13 peaks could be attributed to both flares and/or CMEs affecting the measurements, and the 13 other peaks correlating $\sigma_{\mathrm{phot}}$ could be attributed to another external source.

\subsubsection{Flares and CMEs being able to excite \textit{p}\,modes}

To test whether flares or CMEs could actively excite modes, we examined coincidences between VIRGO/SPMs and GOLF high-energy peaks, from May, 26$^{\mathrm{th}}$ 2003 to May, 03$^{\mathrm{rd}}$, 2022. 
Of the 76 peaks in $\mathcal{S}_{\mathrm{vel}}$, 12 coincided with $\mathcal{S}_{\mathrm{phot}}$. Since both instruments are located on the same satellite, the presence of common high-energy excesses can be attributed to either instrumental factors or the modes themselves. However, only two of these also aligned with $\sigma_{\mathrm{phot}}$ and known flare/CME events, suggesting an external instrumental origin.
The remaining 10 shared peaks likely reflect genuine mode behaviour. Only one aligns with a CME and one with a flare, suggesting a lack of systematic correlation. 
The mode amplitudes vary according to both the instrument. Moreover, this study does not account for the reversal of mode asymmetry observed between intensity and velocity spectra \citep{toutain_1997,toutain_1998,nigam_1998,chaplin_1999}, and the influence of a correlated noise in GOLF data \citep{thiery_2001}.
In particular, Fig.~\ref{fig:spearman_correlation} suggests that if lower-frequency modes are excited simultaneously, the low correlation between instruments could lead to a high-energy peak being detected by one instrument but missed by the other.
Between 2005 and 2009, all peaks detected exclusively in $\mathcal{S}_{\mathrm{vel}}$ correspond to modes in the medium or high-frequency bands, while during the same period, none of the high-energy peaks from the low-frequency band were detected. This observation aligns with the reduced correlation between VIRGO/SPMs and GOLF data for low-frequency modes. Furthermore, none of the 10 peaks correlated with $\mathcal{C}_{\mathrm{phot}}$ but not with $\sigma_{\mathrm{phot}}$ show correlation with $\mathcal{S}_{\mathrm{vel}}$. Given that the convective power level is higher in VIRGO/SPMs data compared to GOLF, these findings reinforce the hypothesis that convection contributes to some of the high-energy peaks observed in $\mathcal{S}_{\mathrm{phot}}$.

Moreover, assuming that flares and CMEs originate only at or near the solar surface, and not within the deeper solar interior, one might expect that if these events were capable of exciting the modes, higher-frequency modes would be preferentially excited before lower-frequency ones. To investigate this, we examined the 82 peaks in $\mathcal{S}_{\mathrm{phot}}$ that show no correlation with $\sigma_{\mathrm{phot}}$. Among these, 23 coincide with CMEs, with 6, 9, and 6 peaks occurring in the low-, medium-, and high-frequency bands, respectively. Similarly, of the 9 peaks associated with X-class flares, 4, 2, and 3 are found in $\mathcal{L}_{\mathrm{phot}}$, $\mathcal{M}_{\mathrm{phot}}$, and $\mathcal{H}_{\mathrm{phot}}$. These distributions indicate that excitation signatures are approximately uniform across frequency bands, inconsistent with a surface-limited excitation mechanism.

Consequently, the hypothesis that these surface events act as a secondary excitation mechanism appears unsupported. This conclusion aligns with the theoretical analysis of \citet{foglizzo_1998}, which argued that flares lack the energy, relative to their spatial extent, to efficiently excite individual modes. Although CME-driven excitation presents a marginally more plausible case, it still depends on assumptions about their spatial scale, inferred from the separation of loop footpoints. Given that flares and CMEs are likely manifestations of the same underlying process, and considering CME onset estimates from \citet{yashiro_2009}, a smaller initiation scale is plausible. In that scenario, the available kinetic energy of CMEs would also be insufficient for effective mode excitation.

\subsubsection{Locating the modes excitation}

Of the 82 peaks in $\mathcal{S}_{\mathrm{phot}}$ not associated with $\sigma_{\mathrm{phot}}$, 44 show no correlation with any of the studied proxies, and 12 correlate only with peaks in $\mathcal{S}{\mathrm{vel}}$. When divided into frequency bands, 20, 23, and 28 of these peaks fall into the low-, medium-, and high-frequency ranges, respectively. Only one peak — observed in December 2012 — is detected across all three frequency bands, but not in $\mathcal{S}{\mathrm{vel}}$. The slight excess of high-frequency events suggests a possible excitation source located in the outermost layers of the convective zone.
Meanwhile, exciting modes simultaneously in the low- and high-frequency bands, but not in the medium one, occurs four times during the total observation period of VIRGO/SPMs. This suggests the presence of a mechanism that primarily acts at the edges of the region where the modes are the most sensitive, rather in the middle layers.
Temporal variations as shown in Fig.~\ref{fig:peaks_over_time_fb} further support the existence of an intricate, time-dependent excitation mechanism, with spatial and temporal characteristics that evolve over the solar cycle. Furthermore, this exciting mechanism is expected to have a time scale lower or similar to the 5-minutes time scale of the modes (details in Appendix~\ref{app:time_scale}).


\section{Conclusions}
\label{sec:conclusion}

In this study, we analysed over 27 years of data from the SoHO VIRGO/SPMs and GOLF instruments using the methodology developed in F+98 to reconstruct the energy time series of low-degree modes ($\ell \leq 2$). 
For each mode with radial order $14 \leq n \leq 25$, we compared the observed energy distribution to the expected distribution under the assumption that the modes are only stochastically excited by convection. 
The general behaviour of the modes is found to be consistent with stochastic excitation. However, similar to previous studies \citep{chaplin_1997, chang_1998}, we observed that the modes are excited at an unusually high rate, indicating non-stochastic characteristics in the excitation mechanisms. 
On the one hand, we computed a proxy for the energy supply rate, which appears to fluctuate over time. The mean $<\dot{E}>$ follows a quasi-biennial oscillation (QBO) pattern, and its variance seems to be anti-correlated with the solar cycle. 
On the other hand, we temporally compared the occurrence of these high-energy peaks with the presence of energy sources, such as X-class flares or highly energetic CMEs. 
Although some high-energy points coincide with these events, there is no direct correlation between the energy released by the event and the occurrence of an energy excess, nor is there a systematic pattern of such events throughout the entire study period. 
By introducing a combination of modes within frequency bands, we aimed to determine the location of the excitation process. Our results suggest the existence of a complex excitation mechanism which is able to excite several modes at once and for which the excitation region changes over time. 

This study did not account for potential deviations from the energy distribution model described by Eq.~\eqref{eq:exponential_distribution} that may arise due to mode asymmetry. Addressing this limitation represents a natural follow-up, potentially through the development of a more comprehensive model that incorporates asymmetric mode profiles. Further extensions could also involve applying the same reconstruction method to higher-degree modes to investigate whether similar correlations exist. This may be feasible using spatially resolved observations, such as those provided by SOI/MDI. Additionally, analogous analyses of mode excitation rates and energy in other solar-like stars would help characterize how energy excess correlates with stellar magnetic activity and its temporal modulation.


\begin{acknowledgements}
      The authors want to acknowledge Leila Bessila, Adam Finley, Kiran Jain, Barbara Perri and Antoine Strugarek for useful comments and discussions.
      The GOLF, VIRGO/SPMs and LASCO instruments on board SoHO are a cooperative effort of many individuals to whom we are indebted. SoHO is a project of international collaboration between ESA and NASA. The authors strongly acknowledge the French space agency, CNES, for its support to GOLF since the launch of SoHO. E.P. and R.A.G. acknowledge the support from the GOLF/SoHO and PLATO Centre National D’Études Spatiales grants.
      S.N.B acknowledges support from PLATO ASI-INAF agreement no. 2022-28-HH.0 "PLATO Fase D".
      The authors want to thank the whole \textit{Kepler} team, all funding councils and agencies that have supported the activities of the KASC and the International Space Science Institute. Funding for the \textit{Kepler} mission is provided by NASA’s Science Mission. We thank the NOAA National Geophysical Data Center's (NGDC) for making their data of the GOES solar flare X-ray freely available in the Solar Data Repository, accessible at \url{ftp://ftp.ngdc.noaa.gov/STP/space-weather/solar-data/solar-features/solar-flares/x-rays/goes/}. We also acknowledge \citet{christensen-dalsgaard_1996} for publicly sharing their solar structure model accessible at \url{https://users-phys.au.dk/jcd/solar_models/}.
      \textit{Softwares}: AstroPy \citep{astropycollaboration_2013,astropycollaboration_2018}, Matplotlib \citep{hunter_2007}, NumPy \citep{vanderwalt_2011}, SciPy \citep{jones_2001}, pandas \citep{mckinney_2010, team_2024} and GYRE \citep{townsend_2013}. Some portions of the text were revised with the help of OpenAI’s ChatGPT (GPT-4), used for improving clarity and scientific language.
\end{acknowledgements}

\bibliographystyle{aa}
\bibliography{Panetier_2025}

\begin{appendix}

\section{Calibration of helioseismic time series}

This section begins by outlining the most recent calibration method applied to the VIRGO/SPMs data, during which outliers caused by instrumental issues are converted into data gaps. As the subsequent analysis is strongly influenced by the contribution of photon noise, we show how the treatment of missing values impacts photon noise, and evaluate the optimal interpolation method for filling these gaps. The photon noise of GOLF over time is also discussed.

\subsection{Calibration of VIRGO/SPMs data}
\label{app:calibration_VIRGO}

The VIRGO/SPMs have three independent photometric channels at 402\,nm (blue), 500\,nm (green) and 862\,nm (red), each consisting of a combination of Si-diode interference filters mounted in a common body which is heated with constant power and always remains a few degrees above the heat sink temperature. This reduces the degradation of optical elements due to condensation of gaseous contaminants. The actual temperature of the detector is monitored by two thermometers and used to correct the sensitivity during data evaluation.

From the beginning of the mission, an issue was identified with the counting electronics, characterised by momentary and random blocking of either the Voltage-to-Frequency Converters (VFCs) or the counters, and resulting in a fixed value output. This phenomenon, referred to as \textit{attractors}, manifests as a tendency for the data to freeze at a certain value. 
The occurrence of attractors can affect multiple channels simultaneously or may be confined to a single channel, depending on the event.
The occurrence rates of attractors from January 23rd, 1996, to October 23rd, 2023, are as follows: 4.1\,\% in the red channel, 5.27\,\% in the green channel, and 8.35\,\% in the blue channel. 
Given their relatively low occurrence percentage, the impact of attractors on long-term time series calculations is minimal, and they were subsequently excluded from the data analysis.

\begin{figure}[!ht]
    \centering
    \includegraphics[width = 0.5\textwidth]{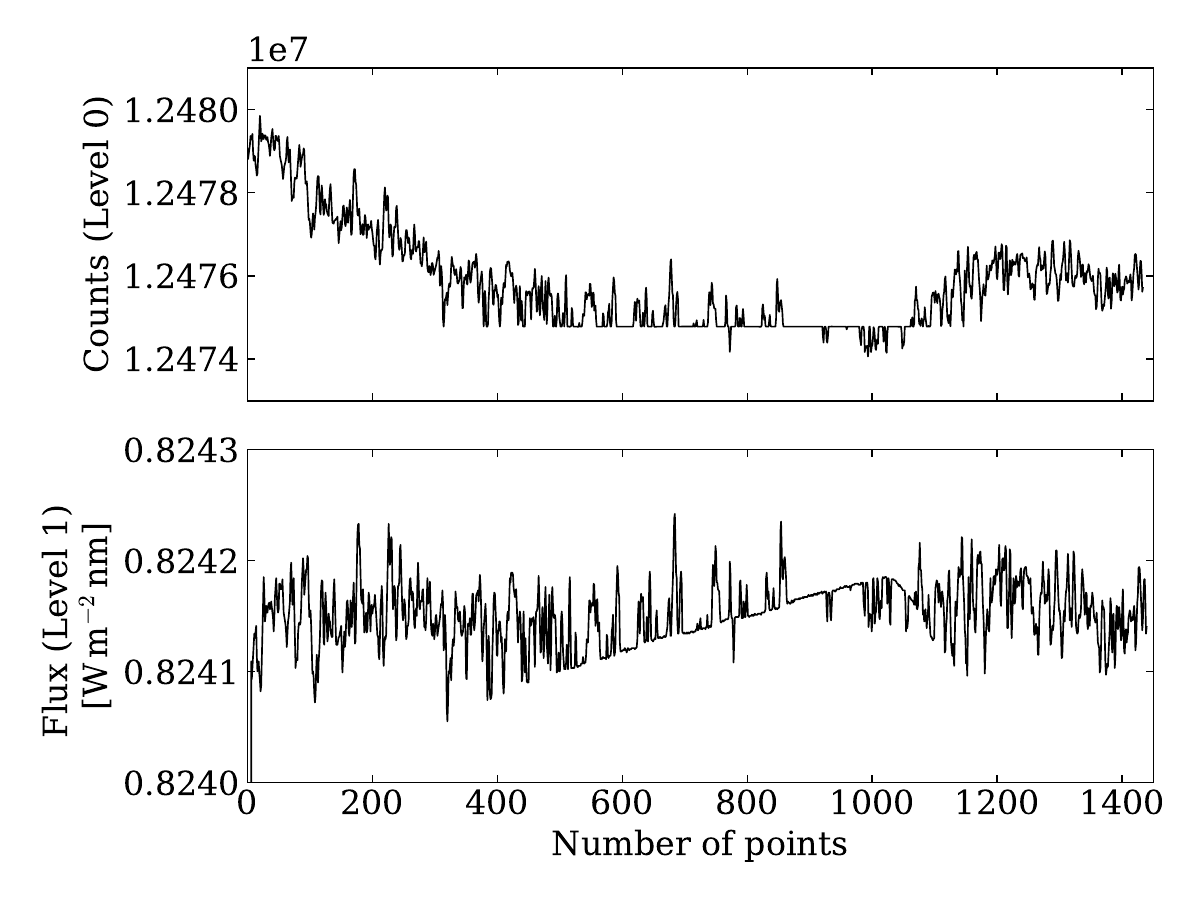}
    \caption{The top panel displays the measured counts (level 0 data), before any corrections. The bottom panel shows the counts transformed into flux (level 1 data), after correction for external effects.}
    \label{fig:attractors}
\end{figure}

\begin{figure}[!ht]
    \centering
    \includegraphics[width = 0.5\textwidth]{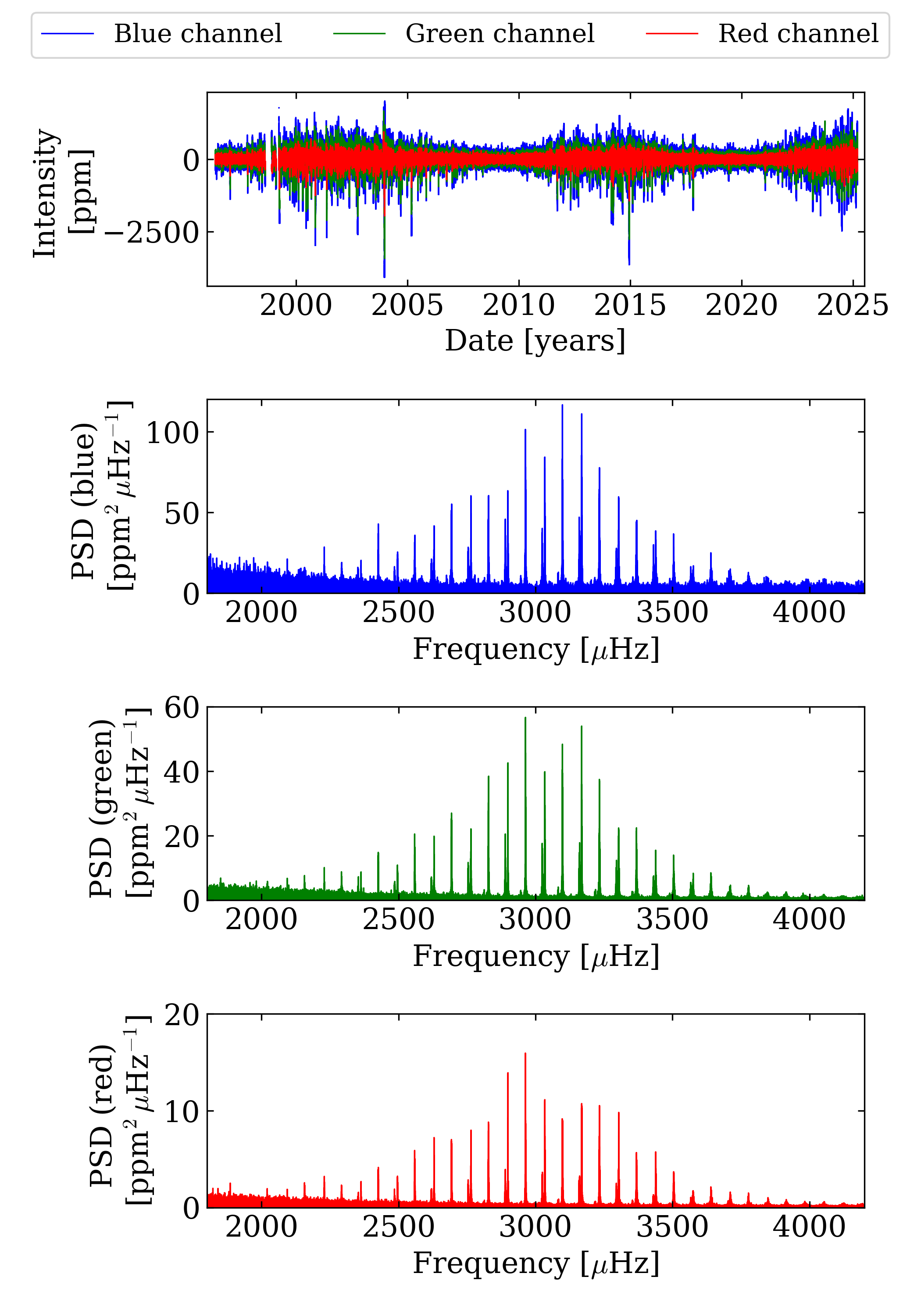}
    \caption{The top panel shows the measured intensity in the three VIRGO/SPMs channels (blue, green and red). The three bottom panels shows the PSD computed from each channel and the x-axis is limited to the \textit{p}-mode frequency range.}
    \label{fig:ts_and_psd}
\end{figure}

Various calibration techniques have been employed since the launch of SoHO to mitigate the effects of this issue \citep{jimenez_2002}. For example, unshared gaps in the data were inferred from signals in other channels in the time series with which \citet{salabert_2017} worked.
However, the most recent data are calibrated as described in the following. The raw counts (level 0 data) are corrected for a range of known factors, including temperature, orbital parameters, pointing, instrument-to-sun distance, relative velocity, and outliers, before being converted into physical quantities (level 1 data).
Since the measured fixed value is no longer constant when corrected, as shown in Fig.~\ref{fig:attractors}, finding an attractor is easier in the counts than in the corrected data. Moreover, in level 0 data, an attractor does not appear exactly as a fixed value but it can change by a couple of counts around a given one. A daily histogram-based technique was then applied to identify all the attractors in level 0 data, by finding the counts that appear more than three times per day. They were replaced by zeros in counts and by missing values in level 1 data, which prevented them from taking part in post-reduction, calibration and analysis. A polynomial fit (7 degrees) and a two-month low-pass filter is then applied to obtain the level 2 time series corrected by degradation. These long-time series have demonstrated a high quality for Helioseismology purposes and magnetic activity evaluation. In Fig.~\ref{fig:ts_and_psd}, the time series of each channel and the corresponding power spectra are shown.

Furthermore, there are also missing values in the data corresponding to the SoHO recovery mission when there were no measurements for about 100 days in 1998, and the upload of a new software in January 1999. 
The asteroseismic proxy used in this work (and described in Sect.~\ref{sec:observations}) is inferred from the mean of red and green channels of VIRGO/SPMS, so that missing values are ignored: for the times when one of both channels has a gap, the value of the other channel is kept. Some gaps are common in both channels, and there are still 2.25\,\% of missing values in this final time series.
After cutting the VIRGO/SPMs asteroseismic proxy into 28 equally long subseries of approximately $376.56\,$days, we interpolated all missing values due to the attractors (gaps smaller than 6 days), by applying a multi-scale discrete cosine transform following inpainting techniques \citep{elad_2005}, which is commonly used for data reduction in asteroseismology \citep[e.g.][]{garcia_2014}. As recommended by \citet{pires_2015}, we performed 100 iterations, for each subseries, in the inpainting procedure. 
43.4\,\% of the missing values are filled by this method. Remaining gaps (larger than 6 days) due to the SoHO recovery mission and the update of the new software were replaced by zeros.

\subsection{Photon noise}

The contribution of photon noise constitutes the dominant component of the signal at frequencies above approximately 8000\,$\mu$Hz \citep{garcia_2019}, allowing for the estimation of the photon noise level in the time series. The VIRGO/SPMs data acquisition system (DAS) operates with a cadence of 3 minutes, corresponding to a fundamental frequency of 5555.55\,$\mu$Hz. The associated Nyquist frequency, defined as half the sampling rate, is 2777.78\,$\mu$Hz. Given the periodic nature of the data sampling, the system introduces harmonics of the fundamental cadence frequency at 5555.55\,$\mu$Hz and an additional harmonic at 8333.33\,$\mu$Hz, which corresponds to a frequency near the Nyquist limit of VIRGO/SPMs. To avoid contamination from these harmonics, particularly the aliasing effect at 8333.33\,$\mu$Hz, the estimation of photon noise is restricted to frequencies below 8200\,$\mu$Hz. This cut-off ensures that the influence of harmonics and aliasing is minimised, thereby providing an accurate representation of the photon noise contribution \citep{salabert_2017}.

\begin{figure}[!ht]
   \centering
   \includegraphics[width = 0.5\textwidth]{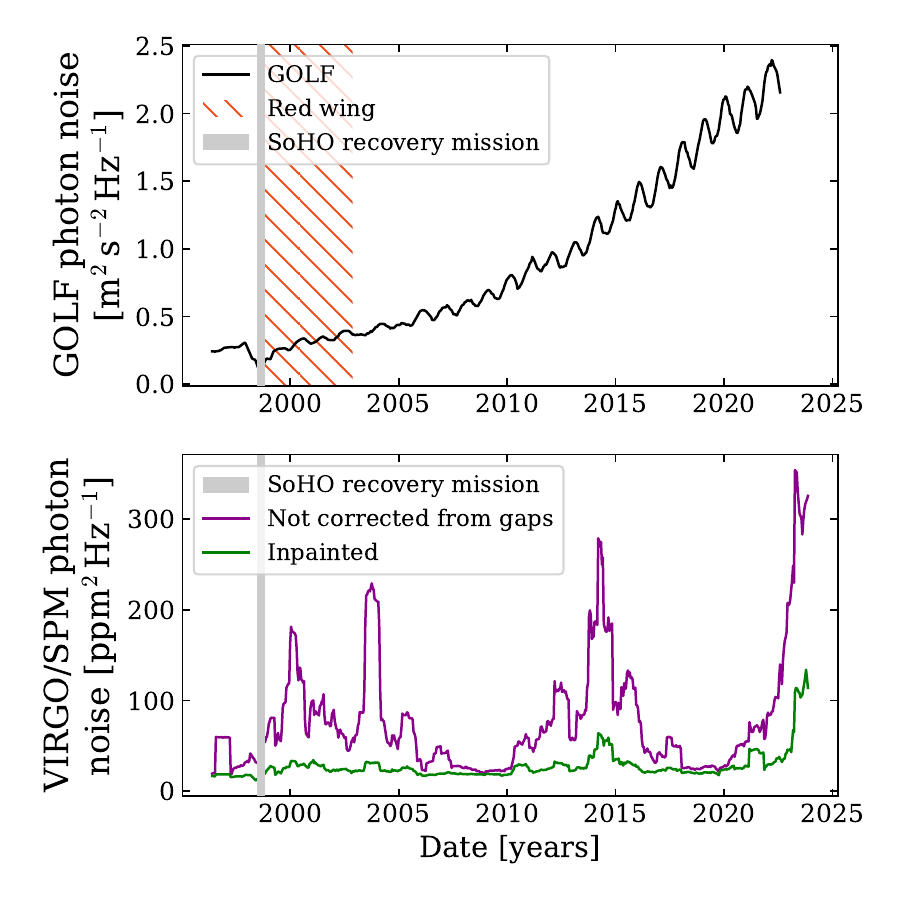}
   \caption{Photon noise estimation from the mean PSD in 8000-8200\,$\mu$Hz range over successive sub-series of 125\,days with 15.625\,days overlap. The top panel displays the photon noise of the GOLF instrument and the bottom panel, the ones from the mean corrected and not-corrected time series of green and red channels of VIRGO/SPMs. The grey zones in each panel indicate the period when SoHO was shut down due to the satellite recovery mission, and the red hatched area in the top panel highlights the dates of the red wing measurements from GOLF.}
              \label{fig:photon_noise}
    \end{figure}
    
For these reasons, to assess the variation in photon noise over the lifetime of the two instruments, estimates were calculated as the mean value of the Power Spectral Density (PSD) in the 8000-8200\,$\mu$Hz range, from 125-day sub-series overlapping by 1/8 the length of the sub-series. 
The variation in photon noise for GOLF is shown in the upper panel of Fig.~\ref{fig:photon_noise}, and the one from the VIRGO/SPMs asteroseismic proxy is in the lower panel.
From this, we see an increase in photon noise of GOLF measurements over time, and a one-year modulation. From the first year of data, to the last one, the mean photon noise per year increased by about 8.6. For VIRGO/SPMs measurements, two different photon noises are shown: the first (green), computed from the time series without correction ($\sigma_{\mathrm{ph,\,nc}}$), where missing values were replaced by zeros; and the second (violet), calculated from the inpainted time series ($\sigma_{\mathrm{ph,\,in}}$). The corresponding mean photon noise is divided by about 2.8 between the two data reductions: 
\[
\overline{\sigma}_{\mathrm{ph,\,nc}}\approx78\,\mathrm{ppm}^2\,\mathrm{Hz}^{-1} \,> \, \overline{\sigma}_{\mathrm{ph,\,in}}\approx28\,\mathrm{ppm}^2\,\mathrm{Hz}^{-1}\,.
\]
The cyclic variation in photon noise from VIRGO/SPMs appears to be linked to the solar magnetic cycle in both cases, with a more significant impact observed during the maximum of Cycle 25. Given that the method used to address gaps in the data heavily affects the photon noise and, consequently, our analysis, we rely solely on the interpolated VIRGO/SPMs dataset for our analysis.

\newpage
\section{The solar radio-flux as a proxy for magnetic cycle}
\label{app:F10.7}

In this section, we describe how the low- and high- magnetic activity epoch were defined based on the solar radio flux index measurements. 

\begin{figure}
\centering
\includegraphics[width=0.5\textwidth,trim={0 0 0 0},clip]{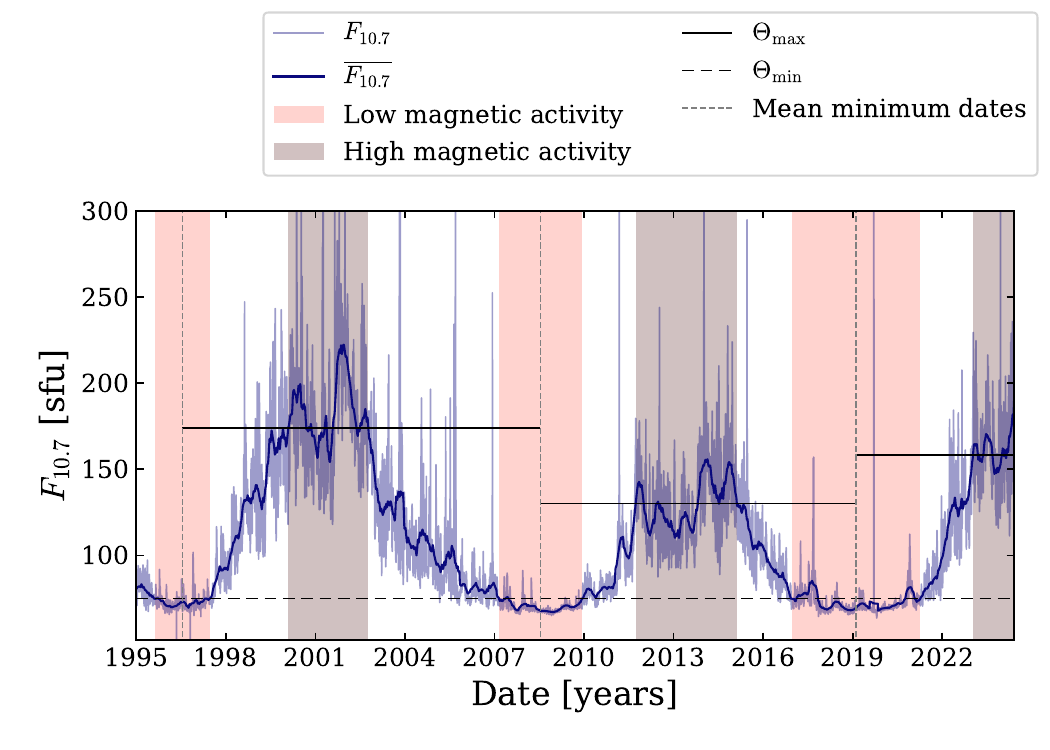}
  \caption{A proxy for the solar magnetic activity: the solar radio-flux given by the $F_{10.7}$ index \citep{tapping_2013} is shown in transparent blue. The dark blue curve shows the mean solar radio-flux filtered with a 101-days window $\overline{F_{10.7}}$. Black lines indicate thresholds (higher than quantiles at 85\,\% for periods of high magnetic activity in continuous lines, and lower than the quantile at 20\,\% for low magnetic activity in dashed line). Epochs used for high magnetic activity dates determination are separated by dashed grey lines that correspond to the mean minimum date of each cycle. Epochs of high and low magnetic activity are highlighted in brown and pink zones and reported in later Figures and in Table~\ref{tab:cycle_extrema}.}
     \label{fig:f10.7}
\end{figure}

The solar radio-flux index \citep[$F_{10.7}$,][]{tapping_2013} represents the total solar radio emission, integrated over one hour, at a wavelength of 10.7\,cm and averaged at a distance of 1\,AU.
Since directly observed measurements of $F_{10.7}$ radio emissions are influenced by variations in the Sun-Earth distance, we opted to use the adjusted flux provided by Natural Resources Canada (NRCan), which accounts for the average distance. 
Complete measurements are publicly available in the NRCan FTP server\footnote{\url{ftp://ftp.seismo.nrcan.gc.ca/spaceweather/solar_flux/daily_flux_values/}}, in solar flux unit (sfu) where $1\,\mathrm{sfu} = 10^{-22}\,\mathrm{W\,m}^{-2}\,\mathrm{Hz}^{-1}$. 
Data are separated into several files because the latest observations (later than October 28$^\mathrm{th}$, 2004) are also directly available from NRCan website\footnote{\url{https://www.spaceweather.gc.ca/forecast-prevision/solar-solaire/solarflux/sx-5-en.php}}. We concatenated two tables to cover the entire SoHO observation period: one containing data from 1996 to 2007, and the second containing latest data from 2004 to today, dropping common values. 
We then calculated the average $\overline{F_{10.7}}$ index, which represents the mean of the solar radio flux index over a 100-day window, shifted for each day. The length of the smoothing window was selected to be sufficiently large, closely matching the 108-day window chosen by \citep{kiefer_2021} for similar smoothing. $F_{10.7}$ and $\overline{F_{10.7}}$ are shown in Fig.~\ref{fig:f10.7}, depicted in light and dark brown, respectively.

\begin{table}
  \caption{Thresholds used to define starting and ending dates of each cycle extremum for Cycles 23, 24 and 25. Dates are listed in Table~\ref{tab:cycle_extrema} and duration of each epoch is in the right column.}
     \label{tab:cycle_extrema_th}
     \centering
     \begin{tabular}{lcc}
        \hline
        \noalign{\smallskip}
        Cycle extremum & $\Theta$ [sfu] & Duration (days) \\
        \noalign{\smallskip}
        \hline
        \noalign{\smallskip}
        Minimum 23 & $<74.92$ & 673  \\
        Maximum 23 & $>173.94$ & 978  \\
        Minimum 24 & $<74.92$ & 1015  \\
        Maximum 24 & $>130.20$ & 1237 \\
        Minimum 25 & $<74.92$ & 1570  \\
        Maximum 25 & $>158.38$ & 550\,\tablefootmark{(a)}  \\
        \noalign{\smallskip}
        \hline
     \end{tabular}
\tablefoot{
    \tablefoottext{a}{Maximum of Cycle 25 is the current phase: we stopped the analysis on July 16th, 2024.}
    }
\end{table}

We defined epochs of magnetic activity extrema from the solar radio flux index, for clarity in the comparisons (see Sect.~\ref{subsec:KS_activity} and \ref{subsec:gamma_dist}). For this, we chose threshold values for the $\overline{F_{10.7}}$ that define the high (and low) periods of magnetic activity. 
For low activity, the threshold $\Theta_\mathrm{min}$ is chosen as the 20\,\% quantile of the $\overline{F_{10.7}}$ values, taken from the starting observation date of VIRGO/SPMs: $\Theta_\mathrm{min}=74.92$\,sfu. Low-activity dates correspond to the extreme dates for each cycle when $\overline{F_{10.7}}$ falls below the threshold.
Figure~\ref{fig:f10.7} highlights these low magnetic activity periods, $\Theta_\mathrm{min}$, and Table~\ref{tab:cycle_extrema_th} reports thresholds values and epochs duration. Figure~\ref{fig:f10.7} illustrates that the durations of low-activity epochs are not all the same: approximately 3 years for minimum of Cycle 24, versus 4 years for minimum of Cycle 24. 
Since epochs of high magnetic activity do not always have the same amplitude, we first define the cycle epochs as the time between two minima, i.e., the time between the successive means of the previously defined low-activity dates.
For instance, the Cycle 23 epoch spans from the average extremum dates of the minimum of Cycle 23 and the average extremum dates of the minimum of Cycle 24, i.e. from 1996-07-18 to 2008-07-16. 
Dashed grey lines in Fig.~\ref{fig:f10.7} separate cycle periods and since we are currently in Cycle 25, the last date of the epoch stops on July 16th, 2024. We then define the thresholds for the high magnetic activity dates as the quantile values at 85\,\% for each cycle period. Thresholds are drawn on Fig.~\ref{fig:f10.7} as well as the periods of high magnetic activity.

\section{Depth of maximum modes sensitivity}
\label{app:kernels}

Acoustic modes are sensitive to different regions of the star, in particular higher-frequency modes are more sensitive to the outer layers of the star. Variations in the magnetic field in some regions should then have a greater influence on the behaviour of modes that are sensitive to the same regions \citep{christensen-dalsgaard_2014}. 
\citet{basu_2012} first computed the sound-speed kernels for several modes grouped by frequency bands, thus determining the sensitivity of these mode groups across the solar radius. When studying the modes $n=11$ to $n=26$, grouped into three frequency bands: 1800-2450\,$\mu$Hz, 2450-3110\,$\mu$Hz, and 3110-3790\,$\mu$Hz, \citet{garcia_2024} showed that their maximum sensitivity occurs in a region located 74\,km to 1575\,km beneath the surface  \citep[using the model of][]{bahcall_2005}. 
In our study, only modes $n=14$ to $n=25$ were considered. To identify the relative location of the non-stochastic excitation of the modes, their energy time series were grouped in sets of 12 and combined into three frequency bands: 2090-2620\,$\mu$Hz, 2620-3160\,$\mu$Hz, and 3160-3710\,$\mu$Hz. 

\begin{figure}[!h]
    \centering
   \includegraphics[width = 0.5\textwidth,trim={0 0 0 0},clip]{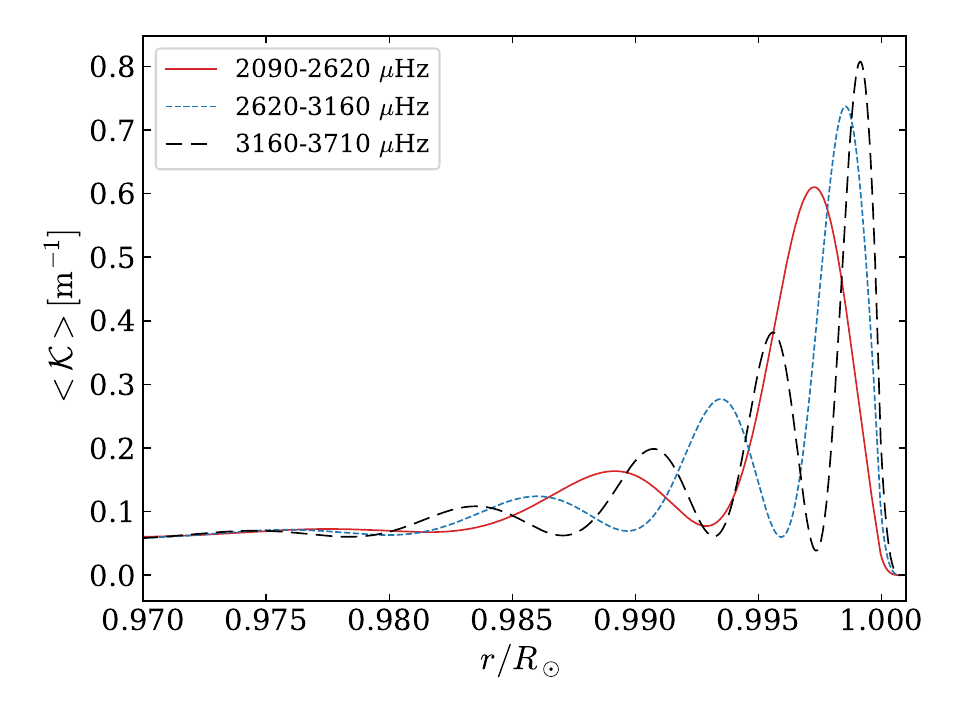}
   \caption{Mean sound-speed kernels $<\mathcal{K}>$ for three frequency bands: 2090-2620\,$\mu$Hz as red continuous line, 2620-3160\,$\mu$Hz as blue dashed line, and 3160-3710\,$\mu$Hz as black long dashed line. For each frequencies band, only $\ell = 0, 1$ and 2 were considered.}
        \label{fig:ss_kernels}
\end{figure}

The mean sound-speed kernels $<\mathcal{K}>$ per frequency range are shown in Fig.~\ref{fig:ss_kernels}.
We briefly remind here the analytic expression of the sound-speed kernels of oscillation modes, as derived by e.g. \citet{gough_1991}. $\mathcal{S}$ is defined as:
\begin{equation}
    \mathcal{S} = \int \left[\xi_r^2 + \ell(\ell+1)\, \xi_h^2 \right] \rho r^2 dr\,,
\end{equation}
where $\xi_r$ and $\xi_h$ are the radial and horizontal components of the displacement $\mathbf{\delta r}$ induced by a small perturbation of the equilibrium state.
The sound-speed kernel $K_{c^2,\rho}(r)$ is thus expressed for each mode ${(n,\ell)}$ as:
\begin{equation}
    \omega^2 \mathcal{S} K_{c^2,\rho}^{(n,\ell)}(r) = \rho c^2 \chi(r)^2 r^2\,,
\end{equation}
where $\omega$ is the angular frequency, $\rho$ the density and $r$ the radius. The adiabatic sound speed $c$ is described as:
\begin{equation}
    c = \sqrt{\frac{\Gamma_1 P}{\rho}}\,,
\end{equation}
where $\Gamma_1=({\partial\ln p}/{\partial\ln \rho})_\mathrm{ad}$ and $P$ is the pressure. 
According to \citet{christensen-dalsgaard_2014}, $\chi(r)$ is the magnitude of $\mathrm{div}(\mathbf{\delta r})$ defined by:
\begin{equation}
    \chi(r) = \frac{1}{r^2}\frac{\mathrm{d(r^2 \xi_r)}}{\mathrm{d}r} - \frac{\ell(\ell+1)}{r} \xi_h\,.
\end{equation}

The oscillation parameters were computed with GYRE \citep{townsend_2013} from the S-model\footnote{\url{https://users-phys.au.dk/jcd/solar_models/}} \citep{christensen-dalsgaard_1996}. The maximum sensitivity for our lower to higher frequency bands are 1895\,km, 1014\,km, and 605\,km beneath the surface, respectively. These results are closer to those found by \citet{basu_2012} than those by \citet{garcia_2024}, as they are highly dependent on the chosen solar structure model.

\section{High-energy peaks detected in the combined time series of VIRGO/SPMs}
\label{app:table_VIRGO_peaks}

The high energy peaks from $\mathcal{S}_{\mathrm{phot}}$ that are detailed in Section~\ref{sec:solutions}, are listed in Table~\ref{tab:peaks_all_comparisons}.

\onecolumn
\begin{longtable}{c|c|c|c|c|c|c|c|c|c}
\caption{List of the dates of the detected high energy peaks in $\mathcal{S}_{\mathrm{phot}}$, and correspondence of the detected peaks in $\sigma_{\mathrm{phot}}$, $\mathcal{C}_{\mathrm{phot}}$ and $\mathcal{S}_{\mathrm{vel}}$, in columns 3 and 4. The time delay is positive if the peak in the photon noise or in GOLF dataset happen after the one in the modes time series. The number of CMEs and X-class flares counted within a window $\delta t = 1.45$\,days around the high energy peaks is indicated in columns 5 and 6. Columns 7 to 9 indicates a cross for peaks that were detected resp. in the low, medium and high frequency ranges.}
\label{tab:peaks_all_comparisons}\\
Date & Hours & $\sigma_{\mathrm{phot}}$ & $\mathcal{C}_{\mathrm{phot}}$ & $\mathcal{S}_{\mathrm{vel}}$ & CMEs & Flares & $\mathcal{L}_{\mathrm{phot}}$ & $\mathcal{M}_{\mathrm{phot}}$ & $\mathcal{H}_{\mathrm{phot}}$ \\
\hline
\endfirsthead
\caption{continued}\\
Date & Hours & $\sigma_{\mathrm{phot}}$ &  $\mathcal{C}_{\mathrm{phot}}$ & $\mathcal{S}_{\mathrm{vel}}$ & CMEs & Flares & $\mathcal{L}_{\mathrm{phot}}$ & $\mathcal{M}_{\mathrm{phot}}$ & $\mathcal{H}_{\mathrm{phot}}$ \\
\hline
\endhead
\hline
\endfoot
\hline
\endlastfoot

1996-09-06 &03:23:20 &   &   &   &   &   &  &  & $\times$ \\
1996-11-26 &03:50:00 & -1.45 & 0 &   & 1 &   &  & $\times$ &  \\
1997-01-27 &08:53:20 &   &   &   &   &   &  &  &  \\
1997-04-29 &23:06:40 &   &   &   &   &   &  &  & $\times$ \\
1997-11-28 &15:16:39 &   &   &   &   & 1 &  &  & $\times$ \\
1998-02-20 &13:09:59 &   &   &   &   &   &  &  &  \\
1998-11-25 &07:49:59 &   & 0 &   & 2 & 3 &  &  &  \\
1998-11-28 &05:16:39 &   &   &   & 2 & 1 &  & $\times$ & $\times$ \\
1998-12-01 &02:43:19 &   &   &   & 1 & 1 &  & $\times$ &  \\
1998-12-04 &00:09:59 &   &   &   &   &   &  & $\times$ &  \\
1998-12-14 &03:13:19 & 0 & 0 &   &   &   & $\times$ & $\times$ &  \\
1999-03-05 &03:39:58 &   &   &   &   &   &  &  & $\times$ \\
1999-03-13 &19:59:58 & 1.45 & 1.45 &   &   &   &  & $\times$ &  \\
1999-09-26 &14:13:18 &   &   &   &   &   &  &  & $\times$ \\
1999-12-05 &00:53:18 &   & 0 &   &   &   &  &  &  \\
2000-05-19 &09:56:38 & 0 & 0 &   & 1 &   & $\times$ &  &  \\
2000-09-23 &17:29:58 & 0 & 0 &   &   &   & $\times$ &  &  \\
2000-11-13 &08:46:38 &   &   &   & 2 &   & $\times$ &  &  \\
2000-12-15 &04:39:58 &   &   &   & 1 &   & $\times$ &  & $\times$ \\
2000-12-25 &07:43:18 & 0 &   &   & 1 &   &  & $\times$ &  \\
2001-04-20 &01:29:58 &   &   &   & 5 &   &  &  &  \\
2001-05-20 &10:39:58 &   &   &   & 1 &   &  & $\times$ &  \\
2001-09-10 &06:59:58 & 1.45 & 1.45 &   & 1 &   & $\times$ &  &  \\
2001-10-04 &21:16:38 & 0 & 0 &   & 1 &   &  &  & $\times$ \\
2002-01-14 &03:49:58 &   &   &   & 2 &   &  &  & $\times$ \\
2002-12-21 &14:16:38 &   & 0 &   & 3 &   &  & $\times$ & $\times$ \\
2003-01-09 &09:39:58 &   & 0 &   & 1 &   & $\times$ & $\times$ &  \\
2003-03-06 &19:49:58 &   &   &   &   &   & $\times$ &  &  \\
2003-05-19 &14:39:58 &   &   &   & 1 &   &  &  & $\times$ \\
2003-10-12 &17:36:38 & 0 & 0 &   &   &   &  &  & $\times$ \\
2003-10-18 &12:29:58 & 0 & 1.45 &  & 2 & 1 & $\times$ &  &  \\
2003-11-16 &10:56:38 &   &   & 0 & 5 &   &  & $\times$ &  \\
2004-02-28 &14:56:38 &   &   &   &   & 1 & $\times$ &  &  \\
2004-05-19 &15:23:18 &   &   &   &   &   &  &  &  \\
2004-12-24 &02:26:38 &   &   &   &   &   &  &  &  \\
2005-03-03 &13:06:38 &   &   & 0 & 2 &   &  & $\times$ &  \\
2005-10-09 &10:53:18 &   &   & 0 &   &   &  & $\times$ &  \\
2005-10-12 &08:19:58 &   &   &   &   &   &  &  &  \\
2006-01-31 &17:56:37 &   &   &   &   &   &  &  &  \\
2006-05-31 &19:53:17 &   & 1.45 &   &   &   & $\times$ &  & $\times$ \\
2007-04-15 &02:46:37 &   & -1.45 &   &   &   &  & $\times$ & $\times$ \\
2007-04-22 &08:23:17 &   &   & 0 &   &   &  &  &  \\
2007-08-07 &09:49:57 &   &   &   &   &   & $\times$ & $\times$ &  \\
2007-10-24 &12:49:57 &   &   &   &   &   & $\times$ & $\times$ &  \\
2008-02-28 &20:23:17 &   &   &   &   &   & $\times$ &  &  \\
2008-06-07 &16:13:17 &   &   & 0 &   &   &  & $\times$ & $\times$ \\
2008-09-03 &22:16:37 &   &   & 0 &   &   &  &  & $\times$ \\
2008-12-27 &05:19:57 &   &   &   & 1 &   &  & $\times$ &  \\
2009-04-29 &04:43:16 &   &   &   &   &   &  &  &  \\
2009-07-19 &05:09:56 &   &   &   & 1 &   &  & $\times$ &  \\
2010-07-18 &19:09:56 & -1.45 & 1.45 &   &   &   &  &  & $\times$ \\
2010-07-27 &11:29:56 &   &   &   &   &   &  &  & $\times$ \\
2010-08-28 &07:23:16 & 1.45 &   &   &   &   &  &  & $\times$ \\
2010-12-06 &03:13:16 &   &   &   &   &   &  &  &  \\
2011-01-27 &05:13:16 &   &   &   &   &   &  & $\times$ &  \\
2011-03-11 &14:53:16 &   &   & 0 &   & 1 & $\times$ &  & $\times$ \\
2011-10-14 &15:13:16 &   &   &   & 1 &   &  &  & $\times$ \\
2012-04-24 &01:16:36 &   &   &   &   &   &  &  & $\times$ \\
2012-07-02 &11:56:35 & 1.45 &   &   & 2 &   &  & $\times$ &  \\
2012-07-25 &15:29:55 &   &   &   & 1 &   &  &  &  \\
2012-12-30 &08:13:15 &   &   &   &   &   & $\times$ & $\times$ & $\times$ \\
2013-02-07 &09:43:15 &   &   &   & 1 &   &  &  &  \\
2013-08-23 &03:56:35 &   &   &   &   &   & $\times$ & $\times$ &  \\
2013-11-17 &23:16:35 &   &   &   & 3 & 1 & $\times$ &  &  \\
2014-01-07 &14:33:15 & 0 & 0 &   & 3 & 1 & $\times$ &  & $\times$ \\
2014-02-05 &12:59:55 & 0 & 0 &   &   &   & $\times$ &  &  \\
2014-05-13 &11:23:15 & 0 & 1.45 &   & 1 &   & $\times$ & $\times$ & $\times$ \\
2014-07-08 &21:33:15 & -1.45 & 0 &   & 3 &   & $\times$ &  & $\times$ \\
2014-09-19 &05:39:55 &   &   &   &   &   &  &  & $\times$ \\
2014-10-26 &20:26:35 & 1.45 & 0 &   &   & 4 &  & $\times$ &  \\
2015-02-12 &08:36:35 &   &   &   & 1 &   & $\times$ &  &  \\
2015-06-02 &07:29:55 &   & 0 &   & 3 &   &  &  &  \\
2016-07-30 &05:06:34 &   &   &   &   &   &  & $\times$ &  \\
2016-08-06 &10:43:14 &   &   & 0 &   &   &  & $\times$ &  \\
2016-12-05 &23:23:14 &   &   &   &   &   &  & $\times$ &  \\
2016-12-26 &05:29:54 &   &   &   &   &   &  &  &  \\
2017-03-20 &03:23:13 &   &   &   &   &   &  &  & $\times$ \\
2017-03-23 &00:49:53 &   & 0 &   &   &   & $\times$ &  & $\times$ \\
2017-05-25 &16:36:33 &   &   &   &   &   &  &  & $\times$ \\
2017-09-06 &20:36:33 & 1.45 & 1.45 &   & 2 & 3 &  & $\times$ & $\times$ \\
2017-09-14 &02:13:13 &   &   &   &   &   &  &  &  \\
2018-01-19 &09:46:33 &   &   & 0 &   &   &  &  &  \\
2018-02-28 &21:59:53 &   &   &   &   &   &  &  &  \\
2018-06-27 &13:13:13 &   &   &   &   &   &  &  & $\times$ \\
2018-07-23 &14:13:13 &   &   &   &   &   &  &  &  \\
2018-07-26 &11:39:53 &   &   &   &   &   &  &  & $\times$ \\
2018-08-04 &03:59:53 &   & 1.45 &   &   &   &  & $\times$ &  \\
2018-11-14 &21:16:33 &   &   &   &   &   &  &  & $\times$ \\
2018-12-29 &17:39:53 &   &   &   &   &   & $\times$ &  &  \\
2019-01-26 &05:23:13 &   &   & 1.45 &   &   &  & $\times$ &  \\
2019-04-05 &16:03:13 &   &   & 0 &   &   &  &  & $\times$ \\
2019-05-16 &04:16:33 & 0 &   &   &   &   &  &  &  \\
2019-05-21 &23:09:53 & -1.45 &   &   &   &   &  &  & $\times$ \\
2019-10-03 &12:19:53 & 0 &   &   &   &   & $\times$ &  &  \\
2019-12-29 &07:39:53 &   &   & 1.45 &   &   &  &  &  \\
2020-01-02 &15:49:53 &   & 1.45 &   &   &   & $\times$ &  &  \\
2020-02-15 &01:29:53 &   &   &   &   &   &  &  &  \\
2020-02-20 &20:23:13 & -1.45 &   &   &   &   &  & $\times$ &  \\
2020-06-02 &13:39:53 &   &   &   &   &   &  & $\times$ &  \\
2020-07-24 &15:39:53 &   &   &   &   &   & $\times$ &  &  \\
2020-08-11 &00:19:53 &   &   &   &   &   &  &  &  \\
2020-08-11 &00:19:53 &   &   &   &   &   &  &  &  \\
2021-06-23 &20:29:53 & 0 & 0 &   &   &   & $\times$ &  &  \\
2021-07-01 &02:06:33 & -1.45 &   &   &   & 1 & $\times$ & $\times$ &  \\
2021-10-30 &14:46:33 &   &   &   & 4 & 1 &  &  &  \\
2022-03-04 &00:53:13 &   &   &   &   &   & $\times$ &  & $\times$ \\
2022-05-21 &03:53:13 & -1.45 &   &   &   &   & $\times$ &  &  \\
2022-12-05 &08:49:53 &   &   &   &   &   &  &  & $\times$ \\
2023-02-20 &01:06:33 &   &   &   & 1 & 1 & $\times$ &  &  \\
\hline
\end{longtable}
\newpage
\twocolumn

\section{Time scale condition for the excitation source}
\label{app:time_scale}

Given that the modes are characterised by a time scale of approximately $T_0=5$\,minutes, we aim to infer a constraint on the time scale of the excitation source required to effectively excite these modes. We assume an oscillation driven by the equation of a forced harmonic oscillator given by:
\begin{equation}
    \ddot{y}+\omega_0^2\,y = f(t)\,,
\end{equation}
where $\omega_0 = 2\pi/T_0$ is the natural frequency of the oscillator and $f(t)$ is the forcing function.
A solution of this is:
\begin{equation}
    y(t) = e^{i\omega_0 t}\int_0^t f(t') \frac{e^{-i\omega_0 t'}}{2i\omega_0} dt' - e^{-i\omega_0 t}\int_0^t f(t') \frac{e^{i\omega_0 t'}}{2i\omega_0} dt'\,.
\label{eq:solution_form}
\end{equation}
Assuming that the forcing function describing the excitation source takes the form:
\begin{equation}
    f(t) =
    \begin{cases}
         a_\mathrm{e}(t) \equiv Y_0\,\omega^2\sin(\omega t)\,, \mathrm{if\,\,} 0<t<\Delta\tau\\
         0\,, 
         \mathrm{elsewhere}
    \end{cases}
\end{equation}
where $\Delta\tau=2\pi/\omega$, $Y_0$ is a constant, and $a_\mathrm{e}(t)$ is the acceleration of the excitation source. The corresponding characteristic speed $v_\mathrm{e}(t)$ is thus:
\begin{equation}
    v_\mathrm{e}(t) \equiv \int_0^t a_\mathrm{e}(t) \mathrm{d}t = Y_0\,\omega[1-\cos(\omega t)]\,,
\end{equation}
and its characteristic amplitude $Y_\mathrm{e}(t)$ is:
\begin{equation}
    Y_\mathrm{e}(t) \equiv \int_0^t v_\mathrm{e} \,\mathrm{d}t = Y_0[\omega t-  \sin(\omega t)]\,,
\end{equation}
where $Y_\mathrm{e}^{\mathrm{max}}=Y_\mathrm{e}(t=\Delta\tau) = 2\pi Y_0$ is the maximum amplitude of the excitation source.

All constant terms apart, the first integral of Eq.~\eqref{eq:solution_form} gives:
\begin{multline}
    \int_0^{\Delta \tau} \sin(\omega t')\,e^{-i\omega_0 t'} \mathrm{d}t' = \int_0^{\Delta \tau} \left(\frac{e^{i\omega t'}-e^{-i\omega t'}}{2i}\right)\,e^{-i\omega_0 t'} \mathrm{d}t' \\
    = \frac{\omega -e^{-i\omega_0 \Delta \tau} [\omega \cos(\omega \Delta \tau) + i\omega_0 \sin(\omega \Delta \tau) ]}{(\omega^2-\omega_0^2)} \,.
\label{eq:4}
\end{multline}
As $\omega\Delta\tau = 2\pi$, this leads to:
\begin{equation}
    \int_0^{\Delta \tau} \sin(\omega t')\,e^{-i\omega_0 t'} \mathrm{d}t' 
    =
    \frac{\omega (1- e^{-i\omega_0 \Delta \tau})}{\omega^2-\omega_0^2}\,.
\end{equation}
In the same way, the second integral of Eq.~\eqref{eq:solution_form} reduces to:
\begin{equation}
    \int_0^{\Delta \tau} \sin(\omega t')\,e^{i\omega_0 t'} \mathrm{d}t' 
    =
    \frac{\omega (1- e^{i\omega_0 \Delta \tau})}{\omega^2-\omega_0^2}\,.
\end{equation}
Thus, when $t>\Delta\tau$, Eq.~\eqref{eq:solution_form}, gives:
\begin{equation}
\begin{split}
    y(t) &= \frac{Y_0\,\omega^3}{2i\omega_0(\omega^2-\omega_0^2)} \left\{ e^{i\omega_0 t} \left(1 -e^{-i\omega_0 \Delta \tau}\right) - e^{-i\omega_0 t} \left(1 -e^{i\omega_0 \Delta \tau}\right) \right\}
    \\
    &= \frac{Y_0\,\omega^3}{2i\omega_0(\omega^2-\omega_0^2)} \bigg\{ 2i\sin(\omega_0t) - 2i \sin[\omega_0(t-\Delta \tau)]\bigg\}\,.
\end{split}
\end{equation}

Let $x\equiv\omega_0/\omega=\Delta\tau/T_0$:
\begin{equation}
\begin{split}
    y(t) &= \frac{Y_0}{x(1-x^2)} \bigg\{\sin(\omega_0t) - \sin\left(\omega_0 t- 2\pi x\right)\bigg\}
    \\
    &= \frac{Y_0}{x(1-x^2)} \bigg\{2\cos(\omega_0t-\pi x) \sin(\pi x)\bigg\}
    \\
    &= \frac{2\pi Y_0}{1-x^2}\, \bigg[\frac{\sin(\pi x)}{\pi x}\bigg] \,\cos(\omega_0t-\pi x)\,.
\end{split}
\end{equation}
Considering $\eta(x)$:
\begin{equation}
    \eta(x) \equiv \frac{1}{1-x^2}\bigg[\frac{\sin(\pi x)}{\pi x}\bigg]\,,
\end{equation}
one can write:
\begin{equation}
    y(t) = Y_\mathrm{e}^\mathrm{max} \,\eta(x) \,\cos(\omega_0t-\pi x)\,.
\end{equation}

\begin{figure}
\centering
\includegraphics[width=0.5\textwidth,trim={0 0 0 0},clip]{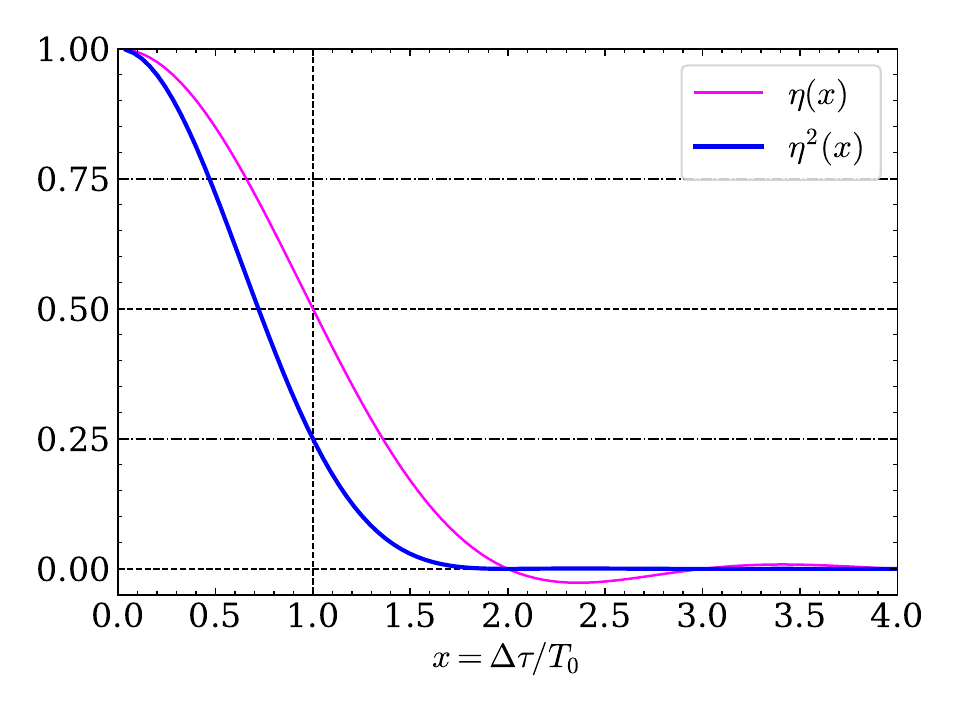}
  \caption{The function $\eta$ (pink line) describes the amplitude of the oscillator relative to the amplitude of the source term for  $t>\Delta\tau$. The blue curve describing $\eta^2$ is a measure of the efficiency of the impulsive excitation.}
     \label{fig:eta_x}
\end{figure}

The function $\eta$ shown in Fig.~\ref{fig:eta_x} describes the fraction of the amplitude of the forcing term that is transferred to the mode amplitude.
The quantity $\eta^2$, also shown in Fig.~\ref{fig:eta_x} can thus be viewed as an efficiency of the energy transfer from the source term to the modes. This efficiency is maximum for $\Delta\tau \ll T_0$. It decreases by approximately 30\,\% when $\Delta\tau= T_0/2$, drops to about
25\,\% when $\Delta\tau = T_0$ and rapidly approaches zero for larger values of the $\Delta\tau/T_0$ ratio.
This calculation illustrates the fact that the impulsive excitation of an oscillator is inefficient if the timescale of the impulsion is larger than the oscillation period.

\end{appendix}

\end{document}